\documentclass[aps,prd,twocolumn,letterpaper,floatfix,showpacs,amsmath]{revtex4} 
\newlength{\picwidth}
 
\usepackage[dvips]{graphicx}
\usepackage{amsmath}

\newcommand{\ern}{\mathcal{E}}
\begin{document}
\renewcommand{\thefigure}{\arabic{figure}}
\title{
Spacetime Encodings IV - The Relationship between Weyl Curvature and Killing Tensors in Stationary Axisymmetric Vacuum Spacetimes.}
\author{Jeandrew Brink}
\affiliation{Theoretical Astrophysics, California Institute of Technology, Pasadena, CA 91103 }

\begin{abstract} 
The problem of obtaining an explicit representation for the fourth invariant of geodesic motion (generalized Carter constant) of an arbitrary stationary axisymmetric vacuum spacetime generated from an Ernst Potential is considered. The coupling between the non-local curvature content of the spacetime as encoded in the Weyl tensor, and the existence of a Killing tensor is explored and a constructive, algebraic test for a fourth order Killing tensor suggested. The approach used exploits the variables defined for the B\"{a}ckland transformations to clarify the relationship between Weyl curvature, constants of geodesic motion,  expressed as Killing tensors, and the solution generation techniques. A new symmetric non-covariant formulation of the Killing equations is given. This formulation transforms the problem of looking for fourth-order Killing tensors in 4D into one of looking for four interlocking two-manifolds admitting fourth-order Killing tensors in 2D.  
\end{abstract}
\pacs{ }

\maketitle

\section{Introduction}
\label{sec:intro}

Very little is understood about the implications that the curvature of a spacetime manifold has for  particle motion within  the spacetime. In the context of Extreme Mass Ratio Inspiral (EMRI) gravitational wave observations careful knowledge of particle motion around compact objects could lead to a spacetime mapping algorithm \cite{JdB0,JdB1}.  This paper provides a framework in which the relationship between orbital invariants and the curvature expressed by the Weyl tensor can be explored.  In particular, it formulates a constructive algebraic test to see whether a particular stationary axisymmetric vacuum (SAV) spacetime  admits an additional invariant, a generalized Carter constant, assumed to result from a  Killing tensor. Only SAV spacetimes that have two commuting Killing vectors $\partial_t$ and $\partial_\phi$ and thus can be generated from a complex Ernst potential are considered.

The existence of a totally symmetric tensor $T_{(\alpha_1\cdots \alpha_m)}$ of order $m$, which obeys the Killing equations $T_{(\alpha_1\cdots \alpha_m;\beta)}=0$ implies that the quantity    
\begin{align}
Q=T^{(\alpha_1\cdots\alpha_m)}p_{\alpha_1} \cdots p_{\alpha_m} \label{Killa}
\end{align}
is constant along a geodesic, and thus provides a constant of motion. In Eq. \eqref{Killa}, $p_\alpha$ indicates the particle momentum.

It was shown in \cite{JdB2} that the condition that a SAV spacetime admits a second-order Killing tensor places direct restrictions on the components of the Weyl tensor. In particular, it limits the Petrov type to  D \cite{Walker,MathTheoryofBlackHoles}.  The approach to the problem of checking whether a particular SAV spacetime admits a Killing tensor of rank $m$ is conceptually simple: all that is required is to formulate the condition that each component of the Killing tensor exists. These integrability conditions result in a number of conditions on the Weyl tensor. Numerical experiments \cite{JdB2,Gair} suggest that a large number of SAV spacetimes may possess a fourth-order Killing tensor. Now, a totally symmetric tensor of order four in four dimensions has 35 possible independent components and the Killing equations impose  56 conditions on the gradients of these components.  In the direct check suggested, the Killing equations have to be satisfied in conjunction with the 10 vacuum field equations. Furthermore, by writing out integrability conditions, many more equations and unknown fields are generated.  The magnitude of the calculation may make the notion of the practical implementation of this idea seem absurd. Possibly for this reason, no literature on and no examples of fourth order Killing tensors in general relativity seems to exist \cite{ExactSolutions,WolfT}. This paper demonstrates how, with some finesse, it is possible to check whether SAV spacetimes admit a fourth-order Killing tensor, and construct its components. 

A number of  ideas lead to the problem becoming tractable. Adopt the point of view that a set of constants of geodesic motion are equivalent to a coordinate system ideally suited to describing the motions of free falling particles in spacetime. Suppose now that an observer with his/her own inertial frame is conducting ballistic tests and keeping track of the velocity and positions of a series of projectiles, in an attempt to discover experimentally what this special coordinate system is. Suppose further that the observer planned to one day  communicate the results of the experiment to someone else traveling through the same SAV spacetime, in an unambiguous manner.  He is concerned about how to orient the coordinate system of the experiment to do so with the greatest ease. A mathematician may suggest that both observer and the traveler orientate their coordinate systems according to the geometry of the spacetime, for instance along the principle null directions of the Weyl tensor, which in most spacetimes will be unique \cite{ExactSolutions}, or by selecting a transverse frame. Whether this advice is experimentally feasible is irrelevant for the rest of the paper, but a well chosen tetrad is introduced in Sec. \ref{SecTetrad}
and all quantities expressed with reference to it.

The investigation \cite{JdB0,JdB1,JdB2} into understanding the relationship between curvature and geodesics was begun to explore the possibility of exploiting the algebraic properties of the solution generation techniques to provide a method of mapping spacetimes and cataloging their properties observationally.  One of the simplest methods of mapping one spacetime onto the next, and one that is directly related to the underlying SL(2,R) symmetry of the SAV field equations, is the discovery of a B\"{a}ckland transformation. Developed by Harrison \cite{HarrisonWET}  and Neugebauer \cite{Neu}, the B\"{a}ckland transformations  (BT) are used to generate a new solution from an existing SAV solution using a function known as a pseudo-potential (or in recent mathematical literature, an ``integral extension'') to carry out the mapping. The choice of variables used in these papers for the curvature quantities makes the field equations particularly transparent and easy to program using algebraic computer systems such as Mathematica. These variables are adopted in this paper. Ultimately, we can ask the question of what additional conditions, if any,  should be imposed on the pseudo-potential so that for a given BT both the curvature components and the  Killing tensor components are mapped to the next solution.

It can be observed that the Killing equations for SAV spacetimes decouple in such a manner that a subgroup of the equations is equivalent to the Killing equations for a two manifold in the ($\rho, z$) plane. This allows us to use the geometric picture, derived from the theory of dynamical systems,  of what a Killing tensor on a two-manifold actually represents \cite{JdB1}, and  to exploit the accompanying symmetries and analytic structure. The remaining Killing equations couple onto the two-dimensional decoupled system in a ``tree-like structure''. Found by inspection, this property allows a large number of Killing tensor components to be sequentially eliminated by well-chosen integrability conditions. The remaining 10 components and their undetermined derivatives are arranged so that repeated differentiation introduces a minimal number of new variables, so the number of integrability conditions grows much more rapidly than the new variables.   Finally the process terminates leading to a spectacularly overdetermined linear system for the components of the Killing tensor and certain of their higher order derivatives, considered to be independent potentials. 

The coefficients of the potentials in the linear system are polynomials in the field variables and their derivatives (the highest derivative of the metric functions that is required is the fifth). Inverting the linear system and writing down the conditions that a solution can be found result in the Killing tensor components, given explicitly as rational functions of the field variables and their derivatives, up to a scaling factor. The consistency conditions for the linear system determine the conditions on the field variables that are required in order for a Killing tensor to exist. The direct inversion of the overdetermined system will be termed the {\it{brute force}} method and is discussed in Sec. \ref{SecKill4}. I argue that it is computationally feasible for existing computers to find the analytic answer. The possible caveat being that once the condition on the field variables is written down, it may not be easy to identify the physical implications of the result.  In practice, 
the main application of the brute force method  would be to construct the Killing tensor components for a given spacetime, once they are known to exist. It should also be noted that the analysis is local, so given a spacetime and enough local derivatives at a point the process described in Sec.~\ref{SecKill4}  would provide an immediate check of whether the Killing equations are consistent at that point. A number of simplifications and insightbuilding special cases are suggested in \ref{SecSimp}, using the general framework of the brute force method.
 
  A more elegant tack than the brute force method is to ask, given a particular SAV spacetime solution that admits a Killing tensor, which of the BT or other solution-generation techniques preserves this property. This question is formulated  clearly mathematically but not answered in Sec. \ref{SecBTF}.

The main effort in the calculation is to arrange quantities so that it does not mushroom out of control.  The actual arrangement can only be understood by reading through the following pages in detail.  The emphasis is on being able to characterize the relationship between curvature, the solution-generation techniques and geodesic motion, so  the r\^{o}le of the Weyl tensor components is made explicit wherever possible. 

In the derivation for second-order Killing tensors given in \cite{JdB2}, it was noted that the problem of seeking a second-order Killing tensor in a 4D spacetime was equivalent to 4 interrelated problems of seeking  second-order Killing tensors on  2D spacetimes.  This allowed the techniques used in two-degree-of-freedom dynamical systems \cite{Hall} to be applied in seeking a solution.  This observation prompted the search for a similar formulation of the 4D, fourth-order Killing equations considered in this paper. An equivalent symmetric formulation of fourth-order Killing equations having a similar property was found and is presented in Sec.~\ref{SecALT}.  In this alternative formulation, the equations are written in the form of the four fourth order Killing equations of a two-manifold with additional interlocking conditions.  This formulation is particularly concise, and appears to offer the possibilty of an explicit analytic solution to the problem. Sec.~\ref{SECCON} concludes the paper with a brief summary of the main results, a few wild speculations, and a number of more serious comments about the outlook of future work. 

\section{SAV Metric and Field Equations}

Any SAV spacetime with two commuting Killing vectors, $\partial_t$ and $\partial_\phi$ can be represented by means of the Lewis-Papapetrou metric,
\begin{align}
ds^2 &=  e^{-2\psi}\left[e^{2\gamma}(d\rho^2+dz^2)+R^2d\phi^2\right]-e^{2\psi}(dt-\omega d\phi)^2. \label{LineEle}
\end{align}
The metric functions can be determined entirely by solutions of the  Ernst equation for the complex potential~$\ern$, 
\begin{align}
\Re (\ern)\ \overline{\nabla} ^2 \ern = \overline{\nabla} \ern \cdot \overline{\nabla}\ern,
\end{align} 
where $\overline{\nabla} ^2= \partial_{\rho\rho}+\frac{1}{\rho}\partial_\rho + \partial_{zz}$ and $\overline{\nabla} = (\partial_\rho,\partial_z)$ . The real part of the Ernst potential is $e^{2\psi} = \Re(\ern)$. Line integrals of the $\ern$~potential determine the functions $\gamma$ and $\omega$. The function $R$ is any harmonic function obeying the equation $R_{,zz}+R_{,\rho\rho} = 0$, and represents a coordinate freedom in this formulation of the metric. One choice that is often used is to set $R=\rho $.

Equivalent to the Ernst formulation of the SAV field equations is a formulation introduced by Harrison \cite{HarrisonWET} and Neugebauer \cite{Neu}. The variables introduced in \cite{HarrisonWET,Neu}   are most suited to performing B\"{a}ckland transformations  and turn out to be directly related to the Ricci rotation coefficients computed for the tetrad, introduced in Sec. \ref{SecTetrad}. It is the Harrison-Neugebauer notation (referred to as $M$ variables) that is adopted for the rest of the paper. This notation makes explicit the relationship between the B\"{a}ckland transformations, the choice of tetrad (Sec. \ref{SecTetrad}), and ultimately the Killing equations (Sec. \ref{SecKill4}), and corresponding constants of motion. Furthermore, the field Eqs. \eqref{FE} expressed in terms of these variables are very easy and efficient to program for symbolic manipulation  (The variable names differ from the original ones used.)

Introduce the complex variables $\zeta = 1/2(\rho + i z)$ and  $\overline{\zeta} = 1/2(\rho - i z)$ and define
\begin{align}
M_1 &= \partial_{\overline{\zeta}} \gamma,&
M_2 &= \frac{\partial_{\overline{\zeta}}   R}{R},&
M_3 &=\frac{\partial_ {\overline{\zeta}}\ern}{\ern+\overline{\ern}},&
M_4 &=\frac{\partial_ {\overline{\zeta}}  \overline{\ern}}{\ern +\overline{\ern}},\notag\\
M_1^* &= \partial_{\zeta} \gamma,&
M_2^* &= \frac{\partial_{\zeta}   R}{R},&
M_3^* &=\frac{\partial_ {\zeta}\ern}{\ern+\overline{\ern}},&
M_4^* &=\frac{\partial_ {\zeta}  \overline{\ern}}{\ern +\overline{\ern}}.\notag\\
\end{align}
Note that the `*' operation is not simply complex conjugation, in fact $\overline{M_1}= M_1^*$,  $\overline{M_2}= M_2^*$, but  $\overline{M_3}= M_4^*$ and  $\overline{M_4}= M_3^*$.

In terms of the  $M$ variables, the field equations can be expressed as
\begin{align}
M_{1,\zeta}&=-\frac{1}{2}(M_3M_4^*+M_4 M_3^*),\notag\\
M_{2,\zeta}&=-M_2M_2^*,\notag\\
M_{3,\zeta}&=-\left(\frac{1}{2}(M_2M_3^*+M_3M_2^*) - M_3M_3^*+M_3M_4^*\right),\notag\\
M_{4,\zeta}&=-\left(\frac{1}{2}(M_2M_4^*+M_4M_2^*) + M_4M_3^*-M_4M_4^* \right), \notag\\
M_{2,\overline{\zeta}}&=-M_2^2+2(M_1 M_2-M_3 M_4). \label{FE}
\end{align}
The remaining  five field equations result from the complex-conjugate expressions of Eqs. \eqref{FE}, which are also required to hold. It should be noted  the field equations determine only certain of the derivatives of the $M$ variables. In particular, $M_{1,\overline{\zeta}}$, $M_{3,\overline{\zeta}}$,  $M_{4,\overline{\zeta}}$, and their complex conjugates are left free and encode the nonlocal content of the spacetime curvature.  These quantities change as one solution is mapped onto another one.  How they enter in the Weyl scalars is shown in Sec.~\ref{SecTetrad}. 

For reference sake, we give the derivatives of the remaining metric components,  
\begin{align}
2\partial_{\overline{\zeta}} \psi &= M_3+M_4,& 
\partial_{\overline{\zeta}} \omega &= R(M_4-M_3)e^{-2\psi}.
\end{align}

\section{Computational Tetrad and Weyl Curvature}
\label{SecTetrad}
In order to understand the interrelationship between Weyl curvature, Petrov Classification, and the Killing equations, it is necessary to introduce a null tetrad. With the particular tetrad constructed below it turns out that the metric functions can be completely eliminated, and the problem written in terms of the $M$ variables and the components of the Killing tensor on the tetrad. 
The first step is to define a local inertial frame or tetrad frame by the basis vectors 
\begin{align}
E_{1}& =-e^{\psi}dt + \omega e^{\psi} d\phi,&E_{2}& =e^{-\psi}R d\phi,\notag\\
 E_{3}& =e^{\gamma-\psi}d\rho,& E_{4}& =e^{\gamma-\psi}dz.
\end{align}
The corresponding contravariant basis vectors are
\begin{align}
E_{1}& =e^{-\psi} \partial_t, & E_{2}& =\frac{\omega e^{\psi}}{R}\partial_t + \frac{e^{\psi}}{R}\partial_\phi ,\notag\\ 
E_{3}&=e^{\psi-\gamma}\partial_\rho, & 
E_{4}& =e^{\psi-\gamma} \partial_z.
\end{align}
Using this inertial frame, define the null tetrad to be
\begin{align}
k &=\frac{1}{\sqrt{2}}(E_{1}+E_{2}),& l &=\frac{1}{\sqrt{2}}( E_{1}-E_{2}),\notag\\
m &=\frac{1}{\sqrt{2}}( E_{3}-iE_{4}),& \overline{m}& =\frac{1}{\sqrt{2}}(E_{3}+iE_{4}).\label{nullTetrad}
\end{align}
The Weyl Tensor coefficients are
\begin{align}
\Psi_0&=C_{\alpha\beta\gamma\delta}k^{\alpha}m^{\beta}k^{\gamma}m^{\delta},\notag\\
\Psi_1&=C_{\alpha\beta\gamma\delta}k^{\alpha}l^{\beta}k^{\gamma}m^{\delta},\notag\\
\Psi_2&=C_{\alpha\beta\gamma\delta}k^{\alpha}m^{\beta}\overline{m}^{\gamma}l^{\delta},\notag\\
\Psi_3&=C_{\alpha\beta\gamma\delta}k^{\alpha}l^{\beta}\overline{m}^{\gamma}l^{\delta},\notag\\
\Psi_4&=C_{\alpha\beta\gamma\delta}\overline{m}^{\alpha}l^{\beta}\overline{m}^{\gamma}l^{\delta}, 
\end{align}
and when expressed on this tetrad in terms of the $Ms$ they become  (setting $V=e^{2\gamma-2\psi}$)
\begin{align}
\Psi_0& = \frac{1}{2V}(- 2M_1^*M_4^* +M_3^*M_4^*+2M_4^{*2}+M_{4,\zeta}^*),\notag\\
\Psi_2& = \frac{1}{4V}(M_2^*M_4+M_2M_4^*-2M_4M_4^*),\notag\\
\Psi_4& =  \frac{1}{2V}(- 2M_1M_4 +M_3M_4+2M_4^{2}+M_{4,\overline{\zeta}}),
\end{align}
and $\Psi_1 = \Psi_3=0$.  Note how the derivatives of the $M$ variables that are not determined by the field equations enter the Weyl scalars $\Psi_4$ and $\Psi_0$. 
The SAV solutions are of Petrov type D if the Weyl scalars expressed on this tetrad obey the identity $\Psi_0\Psi_4 = 9 \Psi_2^2$. In Sec.~\ref{SecKill2} it is shown, as an example of the general formalism how the existence of a second order Killing equation implies that this condition holds.

The Killing equations treated in  Sec. \ref{SecKillT}  require knowledge of the Ricci rotation coefficients  associated with the tetrad  ($k$, $l$, $m$, $\overline{m}$). (See \cite{MathTheoryofBlackHoles} for a definition of rotation coefficients).  Given Eq. \eqref{ROTCO} below, these coefficients serve as dials in the calculation which can be adjusted to generate simpler example problems, as  discussed in Sec.~\ref{SecSimp}. 

In terms of the $M$ variables, the nonzero Ricci rotation coefficients are (recall that the 24 rotation coefficients are antisymmetric in the first two indices $\gamma_{abc}=-\gamma_{bac}$)
\begin{align}
\gamma_{123}&=\frac{M_3^*-M_4^*}{2\sqrt{2V}},&\gamma_{124} &=\frac{M_4-M_3}{2\sqrt{2V}}, \notag\\
\gamma_{131} &=\frac{M_2^*-2M_4^*}{2\sqrt{2V}},&\gamma_{141} &=\frac{M_2-2M_3}{2\sqrt{2V}},\notag\\
 \gamma_{132} &=-\frac{M_2^*}{2\sqrt{2V}},& 
 \gamma_{142} &=-\frac{M_2}{2\sqrt{2V}},\notag\\
\gamma_{231} &=-\frac{M_2^*}{2\sqrt{2V}},&  \gamma_{241} &=-\frac{M_2}{2\sqrt{2V}} , \notag\\
  \gamma_{232} &=\frac{M_2^*-2M_3^*}{2\sqrt{2V}},& \gamma_{242} &=\frac{M_2-2M_4}{2\sqrt{2V}} ,\notag\\
\gamma_{343} &=
-\frac{M_B^*}{2\sqrt{2V}}, & \gamma_{344} &=
\frac{M_B}{2\sqrt{2V}}, \label{ROTCO} \end{align}
where $M_B = \partial_{\overline{\zeta}} \ln V =2M_1-M_3-M_4$.
Vanishing coefficients are
\begin{align}
\gamma_{121}=\gamma_{122}=\gamma_{133}=\gamma_{134}=\gamma_{143}=\gamma_{144}=0,\notag\\
\gamma_{233}=\gamma_{234}=\gamma_{243}=\gamma_{244}=\gamma_{341}=\gamma_{342}=0.
\end{align}
These rotation coefficients clarify the relationship between the $M$ variables used in the BT and the Newman-Penrose tetrad formalism. Furthermore,  if the spacetime is static then $\gamma_{123}=\gamma_{124}=0$ and if the gauge is chosen to be $R=\rho$, $\gamma_{132}=\gamma_{231}=\gamma_{142}=\gamma_{231}$.

Finally, before moving onto the Killing equations, we note that the constant matrix $\eta$ that is used for raising and lowering indices on the null basis has four nonzero components  $\eta^{12}=\eta^{21}=-1$, and $\eta^{34}=\eta^{43}=1$.

\section{Killing Equations}
\label{SecKillT}
On the tetrad basis, the Killing equations for a totally symmetric Killing tensor $T$ of order $m$, or  equivalently for the vanishing of the totally symmetrized intrinsic derivative \mbox{$T_{(a_1\cdots a_m|b)}=0$}, can be expressed in terms of the directional derivatives and rotation coefficients as
\begin{align}
T_{\left(a_1\cdots a_m,b\right)}=m\ \eta^{cd}\gamma_{c(a_1a_2}T_{a_3 \cdots a_m b)d}.
\label{Killtetradgen}
\end{align}
In the case of SAV spacetimes, the absence of functional dependence on the two coordinates $t$ and $\phi$, and the choice of tetrad, Eq. \eqref{nullTetrad}, imply that the directional derivatives in the 1,2 (or $k$, $l$) directions vanish. As a result only directional derivatives in the 3, 4 (or $m$, $\overline{m}$) directions need to be considered. For the tetrad chosen in Eq.~\eqref{nullTetrad}, these derivatives can be expressed as $m^\alpha=\frac{1}{\sqrt{2V}}\partial_\zeta$ and  $\overline{m}^\alpha=\frac{1}{\sqrt{2V}}\partial_{\overline{\zeta}}$.

Both the directional derivatives and the rotation coefficients contain a factor $1/\sqrt{2V}$. When programming the check for the integrability conditions for the Killing components it is convenient to remove this last explicit dependence on the metric functions $V=e^{2\gamma-2\psi}$. To do so we multiply Eq. \eqref{Killtetradgen} by $\sqrt{2V}$ and let $\mathcal{T}$ denote a vector in which all the independent components of the Killing tensor $T$ have been arranged, and we also let $M$ and $M^*$ be four dimensional vectors $M= [M_1,M_2,M_3,M_4]$ and  $M^*= [M_1^*,M_2^*,M_3^*,M_4^*]$.

Then the $n$ Killing equations for a Killing tensor of order $m$ can be represented as the system of equations
\begin{align}
C_A^i \mathcal{T}_{\zeta} + C_B^i \mathcal{T}_{\overline{\zeta}} = MC_D^i \mathcal{T}+M^*C_{E}^i \mathcal{T}, \label{KILLGEN}
\end{align} 
where $i=1, \cdots, n$ and the matrices $C_A^i$, $C_B^i$,  $C_{D}^i$ and  $C_{E^*}^i$ contain only integer entries. This equation makes explicit the coupling between the components of the Killing tensor  and the curvature of the spacetime, expressed in terms of $M$ variables.

The $C^i$ matrices contain a lot of  structure inherited from the Killing equations.  We shall exploit this structure to limit the number of computations that have to be performed to check whether a specific spacetime  admits a Killing tensor of order $m$. In particular, only the cases $m=2$ and $m=4$ have thus far been considered. 


It can always be shown that the Killing equations for the components $T_{(k_1\cdots k_m)}$ with $k_i =\{3,4\}$ decouple from the larger system, and are  the $m$'th order Killing equations for a two-manifold with conformal factor $V$. This observation allows us to use the geometric understanding obtained from the field of dynamical systems about what these Killing tensors represent \cite{JdB1} to identify appropriate variables and possibly useful coordinate transformations.

The remaining components that contain the indices \mbox{$i=\{1,2\}$}  are ultimately coupled to this two-manifold. In the  $m=2$ case, by writing out the appropriate integrability conditions \cite{JdB2} they too can be written in a form of a second-order Killing equation on a two manifold distinct from $V$.  As is shown in Sec. \ref{SecALT}, this property persists for the $m=4$ case, where four two-manifolds admitting a fourth-order Killing tensor are in effect sought. 

If a Killing tensor of even order is considered, it can be shown by explicitly writing out the Killing equations on this tetrad that the equations for  Killing tensor components such as $T_{AABk}$ that admit an odd number of $k_i=\{3,4\}$ and $A,B = \{1,2\}$ indices decouple from the rest, and can  be set to zero, without loss of generality. A more subtle argument using the Jacobi metric \cite{JdB2} and the symmetries of the additional invariant on the two-manifold yields the same result.

In the next two sections, the two examples for $m=2$ and $m=4$ are used to illustrate the ideas described above more concretely. These examples are also used to demonstrate the reductions, in some cases empirical,  that simplify the problem to a computationally manageable size.

\section{Second Order Killing Tensor}
\label{SecKill2}
\begin{table*}[th]
\begin{tabular}{|c|c|c|c|c|c|c|cc|}
\hline
\multicolumn{9}{|c|}{\bf Properties of  Second Order Killing Tensors in SAV Spacetimes}\\ \hline
 \multicolumn{9}{|c|}{$m=2$, $n=10$, K.E. involving  6 Tensor Components}\\ \hline \hline
Column 1  & Column 2   & \multicolumn{2}{c|}{Column 3}& Column 4 & Column 5 & Column 6   & Column 7 &\\  \hline 
\bf Tensor & \bf Number   & \multicolumn{2}{c|}{\bf Derivatives }& \bf Number of ICs & \bf Derivatives & \bf Resulting Unknown   &\bf Number &\\ 
\bf Components & \bf of Comp.&  \multicolumn{2}{c|}{ \bf fixed by K.E.} &  \bf Generated & \bf fixed by IC& \bf Potentials to Determine.  & \bf of Pot.&\\ \hline \hline
$T_{(AB)}$ & 3 &$\ \partial_\zeta\ \cdot\  \ $ & $\partial_{\overline{\zeta}}\ \cdot $&3&-&-&-&\\ \hline 
$T_{34}$ &1 &$\ \partial_\zeta \ \cdot\  \ $ & $\partial_{\overline{\zeta}} \ \cdot$ &1&-&-&- &\\ \hline
$T_{33}$ & 1 &  $\ \partial_\zeta \ \cdot\ \  $ &&-&$\partial_{\overline{\zeta}}^3 \cdot$ &   $\partial_{\overline{\zeta}}^p T_{33}$,   $\ p=\{0,1,2\}$ & 3 & \\ \hline
$T_{44}$ & 1 & &  $\partial_{\overline{\zeta}}\ \cdot$ &-& 
$\partial_{\zeta}^3 \cdot$ &   $\partial_{{\zeta}}^p T_{44}$,   $\ p=\{0,1,2\}$ & 3 & \\ \hline \hline
Total & 6 Comp &  \multicolumn{2}{c|}{10 K.E.} & 4 I.C.&-&-&$L_2= 6$ &\\ \hline
\end{tabular}
\caption{Details of the Second Order Killing Equations (K.E.).  Integrability conditions abbreviated as IC. Indices $k$ and $i$ take on the values ${3,4}$ or equivalently indicate ${m,\overline{m}}$ components.   Indices $A$ and $B$ take on the values ${1,2}$ or  indicate ${l,k}$ components \label{TableKill2}  }
\end{table*}

In the second order ($m=2$) case, there are six non-zero components of the Killing tensor that have to be considered and ten Killing ($n=10$) equations that limit the gradients of these components. A summary of the structure of the Killing tensor components and the gradients determined by the Killing equations is shown in Table \ref{TableKill2}. To avoid any confusion in notation, and to illustrate the general method, the Killing equations are in some cases written out in detail for this example. We begin with the second order version of Eq. \eqref{Killtetradgen},
\begin{align}
\sqrt{2V}T_{\left(ab,f\right)}=2\eta^{nm} \sqrt{2V}\gamma_{n(ab}T_{f)m}.
\label{Killtetrad2}
\end{align}
The 10 Killing Eqs. \eqref{Killtetrad2} can be divided into two groups. 
Four equations for the three Killing tensor components $T_{(k_1k_2)}$,  with  $k_i=\{3,4\}$  decouple from the rest and are the Killing equations for a two-manifold with conformal factor $V$. After setting $M_B = \partial_{\overline{\zeta}}(\ln V)$ and  $M_B^* = \partial_{\zeta}(\ln V)$ these equations are
\begin{align}
T_{33, \zeta} &= M_B^* T_{33},&T_{34, \zeta} &=-\frac{1}{2}(T_{33, \overline{\zeta}}+   M_B T_{33}),  \notag\\
 T_{44, \overline{\zeta}} &= M_B T_{44},& 
T_{34, \overline{\zeta}} &=-\frac{1}{2}(T_{44,\zeta}+ M_B^* T_{44}).\label{SecKillEQ1}
\end{align}
The two equations on the left of \eqref{SecKillEQ1} state that $T_{33}/V$ is an analytic function of $\overline{\zeta}$, and $T_{44}/V$ is an analytic function of $\zeta$   \cite{JdB1}. It is in principle always possible to choose a gauge or $R$-function in \eqref{LineEle} such that two functions are equal to the identity, as was done in \cite{JdB2}.  The equations on the right of \eqref{SecKillEQ1} can be viewed as defining the gradient of  the function $T_{34}$. However, the gradient can only be considered to be valid if the cross-derivatives cancel, or equivalently 
if the integrability condition   
\begin{align}
T_{33, \overline{\zeta\zeta}}+   (M_B T_{33})_{,\overline{\zeta}} = T_{44, \zeta\zeta}+  ( M_B^* T_{44})_{,\zeta} \label{ICEQ2}
\end{align}
holds. If one differentiates Eq. \eqref{ICEQ2} with respect to $\overline{\zeta}$  and substitute in the field Eqs.~\eqref{FE} and Killing Eqs.~\eqref{SecKillEQ1}, then $\partial_{\overline{\zeta}}^3 T_{33}$ can expressed in terms of a linear function of ( $\partial_{\overline{\zeta}}^p T_{33}$  ,  $\partial_{\zeta}^p T_{44}$), where $p=\{0,1,2\}$ and it is considered to be determined by the integrability conditions. This is indicated in the fifth column of Table~\ref{TableKill2}. A similar argument holds for  $\partial_{\zeta}^3 T_{44}$.    The functions $M_B$ and $M_B^*$ and their derivatives $\partial_{\overline{\zeta}}^qM_B$ and $\partial_{\zeta}^qM_B^*$ that are not defined by the field Eqs.~\eqref{FE} always enter as polynomial factors in front of the $T$ variables.

The remaining six Killing equations define the gradients of the functions $T_{11}$, $T_{22}$ and $T_{12}$ (the first row entry on Table \ref{TableKill2}). Setting $\mathcal{T} = [T_{11},T_{12},T_{22},T_{33},T_{34},T_{44}]^t$ and  $\mathcal{S} = [T_{11},T_{12},T_{22}]^t$ these equations have the general form   
\begin{align}
\mathcal{S}^i_{\zeta}= MC_D^i \mathcal{T}+M^*C_{E}^i \mathcal{T},\notag\\
\mathcal{S}^i_{\overline{\zeta}}= M\tilde{C}_D^i \mathcal{T}+M^* \tilde{C}_{E}^i \mathcal{T},
\end{align} 
where $C$ and $\tilde{C}$ are constant matrices with integer entries. The integrability conditions of the function $\mathcal{S}$ are surprisingly simple. They are linear functions in only  $(\partial_{\overline{\zeta}}^p T_{33},\   \partial_{\zeta}^p T_{44})$ where $p=\{0,1\}$ with coefficients polynomial in the $M$-variables (at most quadratic) and the first derivatives of $M_{3\zeta}^*$,  $M_{4\overline{\zeta}}$ (at most linear). No other components of the Killing tensor enter the integrability conditions.

It can be shown \cite{JdB2} that appropriate linear combinations of the integrability conditions for the functions $T_{(AB)}$, where $\{A,B\}$ take on the values $\{1,2\}$, and of Eq. \eqref{ICEQ2} result in three more equations similar to \eqref{ICEQ2}, but for a metric with conformal factor distinct from $V$. This creates the geometric picture that one is looking for four two-manifolds that admit second-order Killing tensors.

By 
writing out the integrability conditions (henceforth IC) for the Killing tensor components whose gradients are fully determined by the Killing equations (the first two row entries in Table~\ref{TableKill2}) and differentiating some of these IC equations to fix the higher order undetermined derivatives of the remaining Killing components (third and fourth entries in Table \ref{TableKill2}, column 5),
one can generate a system of linear equations for the remaining unknown functions  represented in column 6 of Table \ref{TableKill2}.  Taking  additional derivatives of the  ICs (column 4) and substituting in  the field Eqs. \eqref{FE} and  Killing Eqs. \eqref{SecKillEQ1} as well as the expressions for the derivatives fixed by the ICs column 5  further increases the number of equations while keeping the number of unknowns $L_2$ constant.  In doing so we build up a large overdetermined linear system. The number of unknowns, and thus an indication of the size of the linear system is given in column 7 of Table~\ref{TableKill2}. The coefficients in this system are polynomials in the $M$ variables and their derivatives. The field equations and  Killing equations, as written in terms of $M$ variables and expressed on the tetrad presented in this paper, are very easy to program in Mathematica.

The overdetermined system of equations so constructed can be solved for the undetermined Killing tensor components (column 6) up to an overall scaling. (The tensor components that have been eliminated by writing out their ICs can ultimately be constructed from a line integral.) The consistency conditions that ensure that a solution can be found provide polynomial conditions on the $M$ variables and their derivatives.

In the case of the second-order tensor we find, the requirement, amongst others, that  
\begin{align}
0&=\Psi_0\Psi_4-9 \Psi_2^2
\end{align}
or that the SAV spacetime is Petrov type D. The other requirements, namely separability in a certain coordinate system, and the explicit solution of the problem, are discussed more fully in \cite{JdB2}.

The total number in the lower-righthand corner of the tables in many ways represents the largest possible linear system that may be required if the coefficients of the undetermined potentials are to remain polynomial in the $M$ variables, and one is working in a general gauge. A number of simplifications exist that decrease this number and will be mentioned in Sec \ref{SecSimp}. In the case of the second-order Killing equations, we can always work in a gauge  in which the conditions on the $M$ variables can immediately be written down (effectively $L_2 =0$).
A more careful treatment of the second rank Killing equations is given in \cite{JdB2} and the coordinate systems, in which the metric functions are separable, are found and classified there.

The requirement that a second-order Killing tensor exist on a SAV spacetime is very restrictive.  Eq.  \eqref{KILLGEN} makes explicit the coupling between the curvature of a SAV spacetime contained in the $M$ variables on the right hand side and the components of the Killing tensor. One can take the point of view that the $n$ equations of  \eqref{KILLGEN} can be used as a representation of the $M $ variables. With this perspective a rough counting argument immediately implies that it is impossible to represent the eight functionally independent $M$ variables in terms of the six $\mathcal{T}$ variables.
 At the very least, a larger representation, with more free functions in the $\mathcal{T}$ vector, is required, to encompass general spacetimes. It is also not known whether all possible allowed $M$ functions consistent with the field equations can be written in the form  \eqref{KILLGEN}. 
A practical method of checking is proposed in the next section for the fourth-order case.

\section{Fourth-Order Killing Tensors}
\label{SecKill4}
\begin{table*}[th]
\begin{tabular}{|c|c|c|c|c|c|cl|c|}
\hline
\multicolumn{9}{|c|}{\bf Properties of  Fourth Order Killing Tensors in SAV Spacetimes}\\ \hline
 \multicolumn{9}{|c|}{$m=4$, $n=28$, K.E.involving  19 Tensor Components}\\ \hline \hline
Column 1  & Column 2   & \multicolumn{2}{c|}{Column 3}& Column 4 & Column 5 & \multicolumn{2}{c|}{ Column 6 }  & Column 7 \\  \hline 
\bf Tensor & \bf Number   & \multicolumn{2}{c|}{\bf Derivatives }& \bf Number of ICs & \bf Derivatives & \multicolumn{2}{c|}{\bf Resulting Unknown  } &\bf Number \\ 
\bf Components & \bf of Comp.&  \multicolumn{2}{c|}{ \bf fixed by K.E.} &  \bf Generated & \bf fixed by IC&\multicolumn{2}{c|}{ \bf Potentials to Determine. }  & \bf of Pot.\\ \hline \hline
$T_{(ABCD)}$ & 5 &$\ \partial_\zeta\ \cdot\  \ $ & $\partial_{\overline{\zeta}}\ \cdot $&5&-&-&&-\\ \hline 
$T_{(AB34)}$ &3 &$\ \partial_\zeta \ \cdot\  \ $ & $\partial_{\overline{\zeta}} \ \cdot$ &3&-&-&&- \\ \hline
$T_{(3344)}$ &1 &$\ \partial_\zeta \ \cdot\  \ $ & $\partial_{\overline{\zeta}} \ \cdot$ &1&-&-&&- \\ 
 \hline
$T_{(AB33 )}$ & 3 &  $\ \partial_\zeta \ \cdot\ \  $ &&-&$\partial_{\overline{\zeta}}^3 \cdot$ &   $\partial_{\overline{\zeta}}^p T_{(AB33)}$, &  $\ p=\{0,1,2\}$ & 9  \\ \hline
$T_{(3334 )}$ & 1 &  $\ \partial_\zeta \ \cdot\ \  $ &&-&$\partial_{\overline{\zeta}}^3 \cdot$ &   $\partial_{\overline{\zeta}}^p T_{(3334)}$,  & $\ p=\{0,1,2\}$ & 3  \\ \hline
$T_{(3333 )}$ & 1 &  $\ \partial_\zeta \ \cdot\ \  $ &&-&$\partial_{\overline{\zeta}}^5 \cdot$ &   $\partial_{\overline{\zeta}}^p T_{(3333)}$,  & $\ p=\{0,1,2,3,4\}$ & 5  \\ \hline
$T_{(AB44)}$ & 3 & &  $\partial_{\overline{\zeta}}\ \cdot$ &-& 
$\partial_{\zeta}^3 \cdot$ &   $\partial_{{\zeta}}^p T_{(AB44)}$, &  $\ p=\{0,1,2\}$ & 9  \\ \hline
$T_{(3444)}$ & 1 & &  $\partial_{\overline{\zeta}}\ \cdot$ &-& 
$\partial_{\zeta}^3 \cdot$ &   $\partial_{{\zeta}}^p T_{(3444)}$, &  $\ p=\{0,1,2\}$ & 3  \\ \hline
$T_{(4444)}$ & 1 & &  $\partial_{\overline{\zeta}}\ \cdot$ &-& 
$\partial_{\zeta}^5 \cdot$ &   $\partial_{{\zeta}}^p T_{(4444)}$, &  $\ p=\{0,1,2,3,4\}$ & 5  \\ \hline
 \hline
Total & 19 Comp &  \multicolumn{2}{c|}{28 K.E.} & 9 IC&-&-& & $L_4= 34$ \\ \hline
\end{tabular}
\caption{Details of the Fourth Order Killing Equations (K.E.).  Integrability conditions abbreviated as IC. Indices $k$ and $i$ take on the values ${3,4}$ or equivalently indicate ${m,\overline{m}}$ components.   Indices $A$, $B$, $C$, $D$ take on the values ${1,2}$ and indicate ${l,k}$ components \label{TABLE4thorder}}
\end{table*}
The orbital crossing pattern of the numerical integration of the geodesic orbits \cite{JdB1,Gair}, the pole structure of the Ernst equation \cite{ErnstEQ},  the algebraic structure of the Weyl tensor \cite{ExactSolutions}, and the group structure of the B\"{a}ckland transformations \cite{HarrisonWET}, all indicate that an exploration into the existence of a fourth-order Killing tensor on SAV spacetimes is warranted. Possible experimental applications \cite{JdB0} provide urgency.

The analysis is considerably more complicated and computationally expensive  than the second-order case.  It is also highly nonlinear, as opposed to the second-order case which can be linearized  as in \cite{JdB2}. In this section we use the technique outlined in  Sec.~\ref{SecKill2} to make the problem of checking for and constructing a fourth-order Killing tensor on SAV spacetimes, vulnerable to a brute force attack. It is argued  that it lies within the range of current computational capabilities to analytically compute the gradients of the Killing tensor components and the integrability conditions on the $M$ variables representing the curvature  for all SAV spacetimes.  In some sense an answer is assured, although it may be ugly.  In subsequent sections, refinements of this approach are suggested that may yield deeper insight into the field equations, and allow computations of insightful special cases with less computer power.

On SAV spacetimes, the 56 Killing equations for the 35 independent components of a general fourth-order Killing tensor decouple into two groups. As mentioned in Sec.~\ref{SecKillT} the Killing equations for components such as $T_{AABk}$  and $T_{Ak_1k_2k_3}$, with an odd number of $k_i=\{3,4\}$ and $A,B = \{1,2\}$ indices, form an entirely separate group from those with even pairs of indices.  The equations that result from the odd group have more restrictive conditions on the curvature, and are trivially solved by setting them to zero. The more general case involving only the even group is considered.  This reduces the number of tensor components to 19, and the Killing equations by half. Table \ref{TABLE4thorder}  lists the derivative properties of the ``even'' Killing tensor components.

The fourth-order Killing equations under consideration in this Section are written out in full in Appendix \ref{APPENDTETRADIC}.
Three groups of equations can be identified. The decoupled group governing the derivatives $T_{(k_1k_2k_3k_4)}$, with $k_i=\{3,4\}$, are the 6 fourth-order Killing equations for a two-manifold with conformal factor V, Eqs. \eqref{KILLEQF3a} and \eqref{KILLEQF3b}. Two of these Eqs. \eqref{KILLEQF3a}  can be combined to eliminate the $T_{3344}$ component (Table \ref{TABLE4thorder}, third row).  The second group consists of ten Eqs. \eqref{KILLEQF1a}  defining the gradients of the five $T_{(A_1A_2A_3A_4)}$ components where $A_i=\{1,2\}$. These components can be eliminated in favor of five integrability conditions (Table \ref{TABLE4thorder}, first row, and Eq. \eqref{IC1111}). Finally there is a buffer group of 12 Eqs. \eqref{KILLEQF2a}, \eqref{KILLEQF2b} involving the gradients of the mixed terms $T_{A_1 A_2 k_1k_2 }$, six of which (namely Eqs. \eqref{KILLEQF2a}) can be eliminated in favor of the the three integrability conditions for the  $T_{A_1A_2 34}$ terms  (Table \ref{TABLE4thorder}, second row, and Eq. \eqref{IC1133} ). In each case where an integrability condition is written down for a potential mentioned above, it is absent in the resulting IC.

 The ``buffer'' group does not have its counterpart in the second-order Killing equations. Now, a valid solution to the fourth-order Killing equations can be constructed by taking the exterior product of two-second order Killing tensors. This type of solution to the fourth-order Killing equations can be considered to be reducible. For irreducible fourth-order Killing tensors, the ``buffer''  introduces a functional freedom that will necessarily include a greater set of SAV spacetimes than those admitting second-order Killing tensors.  The currently open question is how large this set is.

The nine IC's in column 4 of Table \ref{TABLE4thorder}, (Eqs. \eqref{IC1111} \eqref{IC1133} and \eqref{IC3344} )  and the 34 unknown potentials in column~6 of Table \ref{TABLE4thorder} now form a closed system to differentiation (provided the field Eqs. \eqref{FE}, Killing equations represented in column 3, Table \ref{TABLE4thorder} or Eqs. \eqref{KILLEQF2b} and \eqref{KILLEQF3b}  and the higher order derivatives represented by Column 5 are substituted whenever necessary).  In particular if you consider a system of equations generated by the nine ICs, their first derivatives, all their second derivatives with respect to $\zeta$ and $\overline{\zeta}$, and  the derivatives $\partial_{\zeta\zeta\overline{\zeta}} $IC and  $\partial_{\zeta\overline{\zeta\zeta}} $ IC, a completely overdetermined system for the $L_4=34$ potentials has been constructed. This system may be excessive in that more derivatives of the ICs than what is necessary may have been taken. However it gives one freedom in choosing the pivots during the the elimination of the Killing components as one moves toward finding the conditions on the $M$ variables, so that the answer is assured. The coefficients of the Killing components  are polynomials in the $M$ variables  and their derivatives. At most fourth-order derivatives of the $M$ variables enter these equations.

The elimination process of the Killing components is at this stage {\it{ad hoc}}, guided mainly by inspection and symmetries in the tensor indices.  It is essential to choose linear combinations of the equations such that the pivots, which become denominators in the subsequent manipulations, remain as simple as possible.  Following this process it is possible to eliminate all but the first derivatives of the $T_{3333},\ T_{4444},\ T_{3444},$ and $T_{4333}$ components and the ten basic components themselves (the four remaining $T_{k_1k_2 k_2 k_2}$ and the six remaining $T_{A_1A_2 k k }$ components), before it is no longer apparent how to choose the pivots without introducing complicated denominators and the polynomials in the numerator have become so large that Mathematica has trouble with memory. Thus currently one is in principle one 14-by-14 full-matrix inversion away from an analytic expression for the fourth-order Killing tensor components, in terms of the $M$ variables and their derivatives. The conditions that the $M$ variables have to satisfy can be obtained by factoring the resulting consistency  conditions after the matrix inversion is complete. The polynomials in $M$ and their derivatives are very large. However both the matrix inversion and the subsequent factoring of integrability conditions lie within the range of current computational capacity. They  have to be performed only once to get explicite expressions for the Killing tensor components for all SAV spacetimes.   The only caveat is that once the final factoring step has been performed, the resulting conditions on the $M$ variables may not be in an easily recognizable or compact form.  In the following sections we discuss a series of examples that exploit the framework just given but seek ways to simplify the computations.


\section{Simplification of the General framework} 
\subsection{General considerations}
\label{SecSimp}
Currently the greatest obstacle that hampers the further reduction of the linear system is the complexity of the polynomial coefficients that multiply 
the tensor components. Since the eight $M$ variables and  their derivatives up to the fourth enter  the coefficients, any reduction in the number of $M$ variables is welcome. One somewhat ineffective reduction is the choice of gauge $R=\rho$ which implies that $M_2=M_2^*$. In static spacetimes, we find $M_3 = M_4$, $M_3^* = M_4^*$ and $T_{11kk}=T_{22kk}$,  $T_{1111}=T_{2222}$  and  $T_{1112}=T_{1222}$ by a symmetry argument. This reduces the number of integrability conditions by three, and unknowns by six and  makes the brute force approach more accessible, at the cost of generality. 

The analysis performed in the previous section is local, so the existence of a fourth-order Killing tensor can be disproved by showing that the consistency conditions  fail at a particular point in space. The large overdetermined system generated in the previous section can thus be used to show that there is no consistent solution to the Killing equations given a particular spacetime.

Another simplification that can be imposed is the choice of a gauge in which $T_{4444}/V^2=T_{3333}/V^2=1$, which I shall call the  Killing gauge.  This choice reduces the number of unknown potentials by ten, however it does nothing  to reduce the complexity of the coefficients of the remaining components. Choosing this gauge removes the power of the formalism to check for the existence of a Killing tensor given a metric, but it may be useful in understanding the  structure of the equations. This was the case in second-order Killing equations \cite{JdB2}, where an analogous gauge choice was made.

However the most powerful application of the technique appears to be,  given a metric that is integrable and admits a fourth-order (possibly reducible) invariant, 
and a parameterized family of metrics associated with it, to check which of these metrics retain the property of integrability.
The next subsection suggests an approach that uses this  idea, and that could in principle be used to check the integrability of all SAV metrics. This method also gives more insight into the choice of the $M$ variables.

\subsection{Checking B\"{a}ckland tranformations, a tale of two manifolds}
\label{SecBTF}

This section suggests an elegant method that exploits the framework outlined in Sec \ref{SecKill4} 
and couples it to our knowledge of the solution generation techniques, in order to ascertain which SAV spacetimes admit a fourth-order invariant. It was in large part the elegance of the solution generation techniques, and in particular the B\"{a}ckland transformations (BTs) that originally motivated the choice of the $M$ variables. This choice  makes programming the check proposed in the previous sections computationally possible. The  approach given here can be viewed as an alternative to the brute force computation.

BTs map a valid solution of the SAV field equations onto a another with a different Weyl tensor.  The mapping is carried out with  a so-called pseudo-potential, and for a given starting spacetime  only 16 of these transformations exist \cite{HarrisonWET}. The basic idea is to start with a metric such as Kerr, which is known to admit a fourth-order Killing Tensor, and to ask which of the 16 possible  BTs \cite{HarrisonWET} maintain the property that the new spacetime also admits a  fourth-order Killing tensor. The main advantage of working off an existing solution is that the derivatives of the $M$ functions that are not fixed by the field equations are known and thus the complexity of the polynomials entering the overdetermined linear problem is reduced.

A more mathematical statement of this approach is the following.
Consider two manifolds $({\cal{M}},g)$ and $(\tilde{\cal{M}},\tilde{g})$ with metrics $g$ and $\tilde{g}$, that obey the SAV vacuum field equations.  Suppose that $\cal{M}$ admits a fourth-order Killing tensor and that $\cal{M}$ and $\tilde{\cal{M}}$ are related by  a B\"{a}ckland transformation. Namely,  $\tilde{N}=B N$ and $\tilde{N}^*=B^*N^*$, where $N=\left( M_3,\ M_4,\ M_2\right)$ and $B$ and $B^*$ are the $3\times3$ transformation matrices. The entries of these transformation matrices are very simple rational functions of the pseudo-potentials $q$ and $\lambda$. The 16 possibilities are collated in \cite{HarrisonWET}, where however the notation may vary slightly from that used here. Furthermore $\tilde{M}_1$ is fixed  by the last equation of \eqref{FE}.

The potential $\lambda$ is chosen to be  \mbox{$\lambda = \sqrt{(il-\overline{\zeta})/(il+\zeta)}$}, and as a result 
\begin{align}
d\lambda 
&=\frac{1}{2} \left[ \lambda(\lambda^2-1)M_2^*d\zeta+ \lambda^{-1}(\lambda^2-1)M_2d\overline{\zeta} \right]
\end{align}
The pseudo potential $q$   lies within the prolongation structure and obeys the Ricatti equation \cite{HarrisonWET} 
\begin{align}
dq &= \left[q(1+q\lambda) M^*_3 - (q+\lambda)M^*_4 +\frac{1}{2}(1-q^2)\lambda M^*_2\right]d\zeta
\notag\\
&+ \lambda^{-1} \left[q(q+\lambda)M_4 - (1+q\lambda)M_3 +\frac{1}{2}(1-q^2)M_2)  \right] d\overline{\zeta}.
\label{betaRicatti}
\end{align}
The gradients of $q$ and $\lambda$ are thus determined in terms of the $M$ variables of the known solution. Furthermore, the derivatives of the $M$ variables that are not fixed by the field equations are also explicitely known. An overdetermined linear system of Killing equations such as that suggested in Table \ref{TABLE4thorder} can then be built and used to check the conditions on $q$ that are required to maintain fourth-order integrability.  This approach has not been implemented in full, however the result would  allow us to very accurately quantify the size of  the subgroup of SAV spacetimes that admit a fourth-order Killing tensor and would lead to a deeper understanding about the relationship between the solution generation techniques and the geodesic structure of the resulting spacetimes.

\section{Alternate Formulation of fourth order Killing equations}
\label{SecALT}
The geometric origin of the variables used in the previous sections is obvious by construction: namely, they conform to the  components of the Killing tensor expressed on an orthonormal tetrad basis adapted to the underlying geometry of the metric. In other words, the tetrad is chosen to respect the symmetry imposed by the Killing vectors of the SAV spacetime, and turns out to correspond to the transverse frame of that spacetime. The other spacetime dependent quantities that enter the equations are the rotation coefficients associated with the tetrad.

In this section I present an alternate formulation of the Killing equations, which  was found by inspection.  While the system of equations presented here and those of Appendix \ref{APPENDTETRADIC} are entirely equivalent, the former has  symmetry properties that highlight the difference between static and stationary spacetimes, allowing easy simplification. The formulation presented here has the additional feature that the spacetime-dependent quantities that enter the equations are functions of the metric variables and their derivatives rather than the $M$ variables previously used. The relative simplicity of the resulting equations also allows insight into some of the features of the solutions that are sought. Looking at the structure of these equations led to an ansatz for the Killing tensor components on the equatorial plane of static spacetimes. This ansatz appears to be correct, as shown in \cite{JdBZV2}. The new variables, denoted by $ P_{\text{$<$i:j$>$}}$, are linear combinations of the Killing tensor components on the transverse tetrad and of functions entering the metric.  While there is some vaguely systematic way of searching for the new variables, they are best motivated by the fact they have been shown to work to aid explicite calculation in practice. The original intent was to find a set of variables that cast the Killing equations in as symmetric as possible a form.

Based the derivation of the second-order Killing tensors given in \cite{JdB2}, I suspected that one could find a formulation in which the Killing equations  would take on the form of four ``interlocking'' fourth-order problems for a two manifold.  The formulation given here is the result of that search.  A clearly defined algorithm for finding these coordinates does not exist, except to say grope roughly in this direction and if it is the right way, the answer should have the form suggested in \cite{JdB2} for fourth order Killing tensors on a two manifold, if not continue groping. In what follows I thus resort to the annoying approach of giving the anzatz for the transformation without fully being able to disclose how it was obtained, or what the quantities refer to, except to say that the resulting equations are easier to solve and conform to some intuitive picture I had of their existence before they were found.

All the properties of the Killing equations previously discussed in Secs. \ref{SecKillT} and \ref{SecKill4} are inherited by the new system. The naming convention of the new variables displays some of this structure. For the variable name $ P_{\text{$<$i:j$>$}}$, the index $j$ gives some indication of the differentiability properties of the variable. The designation $j=0$, for example,  indicates that the Killing equations fully determine its gradient, and that it can be removed from this system to yield an integrability condition. There are nine of these variables, $P_{\text{$<$i:0$>$}}$ with $i\in\{-4,\cdots, -1 ,1,\cdots, 5\}$. The indices $j=3$ and $j=4$ indicate that the Killing equations fix derivatives with respect to $\zeta$ and $\overline{\zeta}$ respectively. The index $i$ labels different variables with a fixed derivative structure. The final linear system is built up of the ten variables $ P_{\text{$<$i:j$>$}}$ with $i\in\{1\cdots 5\}$ and $j\in\{3,4\}$, and of higher derivatives.

Without further hesitation, make the following ansatz: define the set of variables that are already decoupled as the fourth-order Killing equation of a two-manifold as
\begin{align}
T_{3444}=& P_{\text{$<$1:4$>$}} e^{2
   \gamma -2 \psi }, & T_{3334}=& P_{\text{$<$1:3$>$}} e^{2 \gamma -2 \psi },\notag\\
T_{3333}=& P_{\text{$<$5:3$>$}} e^{4 \gamma -4 \psi },&T_{4444}=& P_{\text{$<$5:4$>$}} e^{4 \gamma
   -4 \psi },\notag\\ T_{3344}=& P_{\text{$<$-1:0$>$}}.\label{ANZATZ1}
\end{align}
Furthermore, set the mixed components  $T_{ABij}$  equal to

\begin{widetext}
{\allowdisplaybreaks{ 
\begin{align}
T_{1233}=& \frac{1}{12} e^{2 \gamma -4 \psi } \left(-4 e^{2 \psi } P_{\text{$<$1:3$>$}}+e^{4 \psi } \left(-3 \omega ^2 P_{\text{$<$2:3$>$}}+6 \omega 
   P_{\text{$<$3:3$>$}}+6 P_{\text{$<$4:3$>$}}\right)+3 R^2 P_{\text{$<$2:3$>$}}\right),\notag\\
T_{1244}=& \frac{1}{12} e^{2 \gamma -4 \psi } \left(-4 e^{2 \psi }
   P_{\text{$<$1:4$>$}}+e^{4 \psi } \left(-3 \omega ^2 P_{\text{$<$2:4$>$}}+6 \omega  P_{\text{$<$3:4$>$}}+6 P_{\text{$<$4:4$>$}}\right)+3 R^2
   P_{\text{$<$2:4$>$}}\right),\notag\\
T_{1133}=& -\frac{1}{4} e^{2 \gamma -4 \psi } \left(e^{4 \psi } \left(\omega ^2 P_{\text{$<$2:3$>$}}-2 \omega  P_{\text{$<$3:3$>$}}-2
   P_{\text{$<$4:3$>$}}\right)+2 R e^{2 \psi } \left(\omega  P_{\text{$<$2:3$>$}}-P_{\text{$<$3:3$>$}}\right)+R^2 P_{\text{$<$2:3$>$}}\right),\notag
\\T_{1144}=& -\frac{1}{4} e^{2
   \gamma -4 \psi } \left(e^{4 \psi } \left(\omega ^2 P_{\text{$<$2:4$>$}}-2 \omega  P_{\text{$<$3:4$>$}}-2 P_{\text{$<$4:4$>$}}\right)+2 R e^{2 \psi } \left(\omega 
   P_{\text{$<$2:4$>$}}-P_{\text{$<$3:4$>$}}\right)+R^2 P_{\text{$<$2:4$>$}}\right),\notag\\
T_{2233}=& -\frac{1}{4} e^{2 \gamma -4 \psi } \left(e^{4 \psi } \left(\omega ^2
   P_{\text{$<$2:3$>$}}-2 \omega  P_{\text{$<$3:3$>$}}-2 P_{\text{$<$4:3$>$}}\right)-2 R e^{2 \psi } \left(\omega  P_{\text{$<$2:3$>$}}-P_{\text{$<$3:3$>$}}\right)+R^2
   P_{\text{$<$2:3$>$}}\right),\notag\\
T_{2244}=& -\frac{1}{4} e^{2 \gamma -4 \psi } \left(e^{4 \psi } \left(\omega ^2 P_{\text{$<$2:4$>$}}-2 \omega  P_{\text{$<$3:4$>$}}-2
   P_{\text{$<$4:4$>$}}\right)-2 R e^{2 \psi } \left(\omega  P_{\text{$<$2:4$>$}}-P_{\text{$<$3:4$>$}}\right)+R^2 P_{\text{$<$2:4$>$}}\right),\notag\\
T_{1234}=& \frac{1}{8} e^{-2
   \psi } \left(-4 e^{2 \psi } P_{\text{$<$-1:0$>$}}+e^{4 \psi } \left(\omega  \left(2 P_{\text{$<$-3:0$>$}}-\omega  P_{\text{$<$-2:0$>$}}\right)+2
   P_{\text{$<$-4:0$>$}}\right)+R^2 P_{\text{$<$-2:0$>$}}\right),\notag\\
T_{1134}=& \frac{1}{8} e^{-2 \psi } \left(e^{4 \psi } \left(\omega  \left(2 P_{\text{$<$-3:0$>$}}-\omega 
   P_{\text{$<$-2:0$>$}}\right)+2 P_{\text{$<$-4:0$>$}}\right)+2 R e^{2 \psi } \left(P_{\text{$<$-3:0$>$}}-\omega  P_{\text{$<$-2:0$>$}}\right)+R^2
   \left(-P_{\text{$<$-2:0$>$}}\right)\right),\notag\\
T_{2234}=& \frac{1}{8} e^{-2 \psi } \left(e^{4 \psi } \left(\omega  \left(2 P_{\text{$<$-3:0$>$}}-\omega 
   P_{\text{$<$-2:0$>$}}\right)+2 P_{\text{$<$-4:0$>$}}\right)-2 R e^{2 \psi } \left(P_{\text{$<$-3:0$>$}}-\omega  P_{\text{$<$-2:0$>$}}\right)+R^2
   \left(-P_{\text{$<$-2:0$>$}}\right)\right).\label{ANZATZ2}
\end{align}}}
Finally, set the components of the form $T_{(A_1A_2A_3A_4)}$, $A_i=\{1,2\}$ equal to
{\allowdisplaybreaks{  \begin{align}
T_{1111}=&\frac{9}{64}\left[ e^{-4 \psi } P_{\text{$<$1:0$>$}} \left(R+e^{2 \psi } \omega \right)^4+4 e^{-2 \psi } P_{\text{$<$2:0$>$}} \left(R+e^{2 \psi } \omega
   \right)^3+8 P_{\text{$<$3:0$>$}} \left(R+e^{2 \psi } \omega \right)^2\right]\notag\\
&+\frac{9}{16}\left[ e^{2 \psi } P_{\text{$<$4:0$>$}} \left(R+e^{2 \psi } \omega
   \right)+ e^{4 \psi } P_{\text{$<$5:0$>$}}\right],\notag
\\
T_{2222}=&\frac{9}{64}\left[ e^{-4 \psi } P_{\text{$<$1:0$>$}} \left(R-e^{2 \psi } \omega \right)^4+4 e^{-2 \psi
   } P_{\text{$<$2:0$>$}} \left(e^{2 \psi } \omega -R\right)^3+8 P_{\text{$<$3:0$>$}} \left(R-e^{2 \psi } \omega \right)^2 \right]
\notag\\&+\frac{9}{16}\left[ e^{2 \psi }
   P_{\text{$<$4:0$>$}} \left(e^{2 \psi } \omega -R\right)+ e^{4 \psi } P_{\text{$<$5:0$>$}}\right],\notag\\
T_{1112}=&\left[\frac{9}{64} e^{-4 \psi } P_{\text{$<$1:0$>$}} \left(e^{4
   \psi } \omega^2 -R^2\right) +\frac{3}{8} e^{-2 \psi } P_{\text{$<$-2:0$>$}} -\frac{9}{32} e^{-2 \psi }
   P_{\text{$<$2:0$>$}} \left(R-2 e^{2 \psi } \omega \right)\right] \left(R+e^{2 \psi } \omega \right)^2
\notag\\
&+\left(\frac{9}{8} e^{2 \psi } \omega  P_{\text{$<$3:0$>$}} 
-\frac{3}{4} P_{\text{$<$-3:0$>$}}\right) \left(R+e^{2 \psi } \omega
   \right)-\frac{3}{4} e^{2 \psi } P_{\text{$<$-4:0$>$}}+\frac{9}{32} e^{2 \psi }
   P_{\text{$<$4:0$>$}} \left(R+2 e^{2 \psi } \omega \right)+\frac{9}{16} e^{4 \psi } P_{\text{$<$5:0$>$}},\notag\\
T_{1222}=&\left[\frac{9}{64} e^{-4 \psi } P_{\text{$<$1:0$>$}}
    \left(e^{4 \psi } \omega^2 -R^2\right)+\frac{3}{8} e^{-2 \psi } P_{\text{$<$-2:0$>$}} +\frac{9}{32}
   e^{-2 \psi } P_{\text{$<$2:0$>$}} \left(R+2 e^{2 \psi } \omega \right) \right]  \left(R-e^{2 \psi } \omega \right)^2  \notag\\
&+\left( \frac{9}{8} e^{2 \psi } \omega  P_{\text{$<$3:0$>$}}   -\frac{3}{4} P_{\text{$<$-3:0$>$}}\right)    \left(e^{2 \psi } \omega -R\right)-\frac{3}{4} e^{2 \psi } P_{\text{$<$-4:0$>$}}+\frac{9}{32} e^{2 \psi }
   P_{\text{$<$4:0$>$}} \left(2 e^{2 \psi } \omega -R\right)+\frac{9}{16} e^{4 \psi } P_{\text{$<$5:0$>$}},\notag\\
T_{1122}=&P_{\text{$<$-1:0$>$}}+\frac{9}{64} e^{-4 \psi }
   P_{\text{$<$1:0$>$}} \left(R^2-e^{4 \psi } \omega ^2\right)^2+\left(\frac{1}{2} e^{-2 \psi } P_{\text{$<$-2:0$>$}} +\frac{9}{16} \omega 
   P_{\text{$<$2:0$>$}} \right)  \left(e^{4 \psi } \omega ^2-R^2\right)\notag\\
&-e^{2 \psi } \omega  P_{\text{$<$-3:0$>$}}-\frac{3}{8} P_{\text{$<$3:0$>$}} \left(R^2-3 e^{4 \psi } \omega
   ^2\right)-e^{2 \psi } P_{\text{$<$-4:0$>$}}+\frac{9}{16} e^{4 \psi } \omega  P_{\text{$<$4:0$>$}}+\frac{9}{16} e^{4 \psi } P_{\text{$<$5:0$>$}}.
\label{ANZATZ3}\end{align}}}
\end{widetext}
After this horror of an ansatz, the resulting Killing equations are  much more amicable and concise. 
Substituting the ansatz in Eqs. \eqref{ANZATZ1}, \eqref{ANZATZ2}, and \eqref{ANZATZ3}, into the Killing Eqs. \eqref{KILLEQF1a}, \eqref{KILLEQF2a}, \eqref{KILLEQF2b}, \eqref{KILLEQF3a} and \eqref{KILLEQF3b} yields, after simplification, the following equivalent system of  Killing equations; the terms whose gradients are not fully defined and that enter the linear system used to construct the integrability conditions are given below,
\begin{align}
P_{<i:3>,\zeta }&= -f_{i,\overline{\zeta}} P_{<5:3>}-\frac{1}{4}f_i P_{\text{$<$5:3$>$},\overline{\zeta}}& i\in\{1 \cdots 4\},\notag\\
P_{<i:4>,\overline{\zeta}}&= -f_{i,\zeta } P_{<5:4>}-\frac{1}{4}f_i P_{\text{$<$5:4$>$,$\zeta $}}& i\in\{1 \cdots 4\},\notag\\
P_{\text{$<$5:3$>$,$\zeta $}}&=   P_{\text{$<$5:4$>$,$\overline{\zeta}$}} = 0. 
\label{PKILL3}
\end{align}
 The functions $f_i$  entering these equations are defined in terms of the metric functions as 
{\allowdisplaybreaks{
\begin{align}
f_1&= e^{2 \gamma -2 \psi }=V,&f_2&= \frac{2 e^{2 \gamma }}{3 R^2},\notag\\
f_3&= \frac{2 e^{2 \gamma } \omega }{3 R^2},& f_4&= \frac{e^{2 \gamma } \left(R^2 e^{-4 \psi }-\omega
   ^2\right)}{3 R^2}.
\end{align}}}
Note that in Eqs. \eqref{PKILL3} $P_{\text{$<$5:4$>$}}$ is an analytic function of $\zeta$ and indicates a gauge freedom still present in the metric. Without loss of generality one can set $P_{\text{$<$5:4$>$}} =P_{\text{$<$5:3$>$}}=1$, with vanishing derivatives further simplifying the expressions \eqref{PKILL3} to 
\begin{align}
P_{<i:3>,\zeta }&= -f_{i,\overline{\zeta}},& P_{<i:4>,\overline{\zeta}}&= -f_{i,\zeta }& i\in\{1 \cdots 4\}.
\label{PKILL3b}
\end{align}

The equations governing the field components whose gradients are fully described are now given by 
{\allowdisplaybreaks{
\begin{align}
P_{\text{$<$-1:0$>$,$\zeta $}}=& -\frac{4}{3} P_{\text{$<$1:3$>$}} f_{\text{1,$\overline{\zeta}$}}-\frac{2}{3} f_1
   P_{\text{$<$1:3$>$,$\overline{\zeta}$}},\notag\\
P_{\text{$<$-1:0$>$,$\overline{\zeta}$}}=& -\frac{4}{3} P_{\text{$<$1:4$>$}}
   f_{\text{1,$\zeta $}}-\frac{2}{3} f_1 P_{\text{$<$1:4$>$,$\zeta $}},\notag\\
P_{\text{$<$-2:0$>$,$\zeta $}}=& -2 P_{\text{$<$1:3$>$}} f_{\text{2,$\overline{\zeta}$}}-f_2 P_{\text{$<$1:3$>$,$\overline{\zeta}$}}\notag\\
&-2   f_{\text{1,$\overline{\zeta}$}} P_{\text{$<$2:3$>$}}-f_1 P_{\text{$<$2:3$>$,$\overline{\zeta}$}},\notag\\
P_{\text{$<$-2:0$>$,$\overline{\zeta}$}}=& -2 P_{\text{$<$1:4$>$}} f_{\text{2,$\zeta $}}-f_2
   P_{\text{$<$1:4$>$,$\zeta $}}\notag\\
& -2 f_{\text{1,$\zeta $}} P_{\text{$<$2:4$>$}}-f_1 P_{\text{$<$2:4$>$,$\zeta $}},\notag\\
P_{\text{$<$-3:0$>$,$\zeta $}}=& -2 P_{\text{$<$1:3$>$}}
   f_{\text{3,$\overline{\zeta}$}}-f_3 P_{\text{$<$1:3$>$,$\overline{\zeta}$}}\notag\\
& -2 f_{\text{1,$\overline{\zeta}$}} P_{\text{$<$3:3$>$}}-f_1
   P_{\text{$<$3:3$>$,$\overline{\zeta}$}},\notag\\
P_{\text{$<$-3:0$>$,$\overline{\zeta}$}}=& -2 P_{\text{$<$1:4$>$}}
   f_{\text{3,$\zeta $}}-f_3 P_{\text{$<$1:4$>$,$\zeta $}}\notag\\
&-2 f_{\text{1,$\zeta $}} P_{\text{$<$3:4$>$}}-f_1 P_{\text{$<$3:4$>$,$\zeta   $}},\notag\\
P_{\text{$<$-4:0$>$,$\zeta $}}=& -2 P_{\text{$<$1:3$>$}} f_{\text{4,$\overline{\zeta}$}}-f_4 P_{\text{$<$1:3$>$,$\overline{\zeta}$}}\notag\\
&-2
   f_{\text{1,$\overline{\zeta}$}} P_{\text{$<$4:3$>$}}-f_1 P_{\text{$<$4:3$>$,$\overline{\zeta}$}},\notag\\
P_{\text{$<$-4:0$>$,$\overline{\zeta}$}}=& -2 P_{\text{$<$1:4$>$}} f_{\text{4,$\zeta $}}-f_4 P_{\text{$<$1:4$>$,$\zeta $}}\notag\\
&-2 f_{\text{1,$\zeta $}} P_{\text{$<$4:4$>$}}-f_1   P_{\text{$<$4:4$>$,$\zeta $}},\notag\\
P_{\text{$<$1:0$>$,$\zeta $}}=& -8 P_{\text{$<$2:3$>$}} f_{\text{2,$\overline{\zeta}$}}-4 f_2 P_{\text{$<$2:3$>$,$\overline{\zeta}$}},\notag\\
P_{\text{$<$1:0$>$,$\overline{\zeta}$}}=& -8 P_{\text{$<$2:4$>$}} f_{\text{2,$\zeta $}}-4 f_2 P_{\text{$<$2:4$>$,$\zeta
   $}},\notag\\
P_{\text{$<$2:0$>$,$\zeta $}}=&+ 4  P_{\text{$<$2:3$>$}} f_{\text{3,$\overline{\zeta}$}}+2 f_3 P_{\text{$<$2:3$>$,$\overline{\zeta}$}}\notag\\
&+4 f_{\text{2,$\overline{\zeta}$}} P_{\text{$<$3:3$>$}}+2 f_2
   P_{\text{$<$3:3$>$,$\overline{\zeta}$}},\notag\\
P_{\text{$<$2:0$>$,$\overline{\zeta}$}}=& +4 P_{\text{$<$2:4$>$}} f_{\text{3,$\zeta $}}+2 f_3 P_{\text{$<$2:4$>$,$\zeta $}}
\notag\\
&+4 f_{\text{2,$\zeta $}} P_{\text{$<$3:4$>$}}+2
   f_2 P_{\text{$<$3:4$>$,$\zeta $}},\notag\\
P_{\text{$<$3:0$>$,$\zeta $}}=&+ 2 P_{\text{$<$2:3$>$}} f_{\text{4,$\overline{\zeta}$}}+f_4 P_{\text{$<$2:3$>$,$\overline{\zeta}$}}+2
   f_{\text{2,$\overline{\zeta}$}} P_{\text{$<$4:3$>$}}\notag\\
&-4 P_{\text{$<$3:3$>$}} f_{\text{3,$\overline{\zeta}$}}-2 f_3 P_{\text{$<$3:3$>$,$\overline{\zeta}$}}+f_2
   P_{\text{$<$4:3$>$,$\overline{\zeta}$}},\notag\\
P_{\text{$<$3:0$>$,$\overline{\zeta}$}}=&+ 2 P_{\text{$<$2:4$>$}} f_{\text{4,$\zeta $}}+f_4 P_{\text{$<$2:4$>$,$\zeta $}}+2 f_{\text{2,$\zeta
   $}} P_{\text{$<$4:4$>$}}\notag\\
&-4 P_{\text{$<$3:4$>$}} f_{\text{3,$\zeta $}}-2 f_3 P_{\text{$<$3:4$>$,$\zeta $}}+f_2 P_{\text{$<$4:4$>$,$\zeta
   $}},\notag\\
P_{\text{$<$4:0$>$,$\zeta $}}=& -8 P_{\text{$<$3:3$>$}} f_{\text{4,$\overline{\zeta}$}}-4 f_4 P_{\text{$<$3:3$>$,$\overline{\zeta}$}}\notag\\
&  -8
   f_{\text{3,$\overline{\zeta}$}} P_{\text{$<$4:3$>$}}-4 f_3 P_{\text{$<$4:3$>$,$\overline{\zeta}$}},\notag\\
P_{\text{$<$4:0$>$,$\overline{\zeta}$}}=& -8 P_{\text{$<$3:4$>$}} f_{\text{4,$\zeta $}}-4 f_4 P_{\text{$<$3:4$>$,$\zeta $}}\notag\\&-8 f_{\text{3,$\zeta $}} P_{\text{$<$4:4$>$}}-4
   f_3 P_{\text{$<$4:4$>$,$\zeta $}},\notag\\
P_{\text{$<$5:0$>$,$\zeta $}}=& -8 P_{\text{$<$4:3$>$}} f_{\text{4,$\overline{\zeta}$}}-4
   f_4 P_{\text{$<$4:3$>$,$\overline{\zeta}$}},\notag\\
P_{\text{$<$5:0$>$,$\overline{\zeta}$}}=& -8 P_{\text{$<$4:4$>$}} f_{\text{4,$\zeta $}}-4 f_4 P_{\text{$<$4:4$>$,$\zeta $}}\ .\label{KILLPV2}
\end{align}}}
An additional advantage of this formulation is that the terms on the right-hand side of the gradients do not contain any of the $ P_{<i:0>}$ variables whose gradients are being defined. Thus, unlike the case treated in App. \ref{APPENDTETRADIC}, it is immediately obvious that the ICs contain only the ten variables  $ P_{<i:3>}$ and   $ P_{<i:4>}$.

In the event that the spacetime is static rather than stationary, $f_3 = 0$, and the following $P_{<i,j>}$ variables can be set to zero:
\begin{align}P_{\text{$<$3:3$>$}}&=P_{\text{$<$3:4$>$}}=P_{\text{$<$2:0$>$}}=
P_{\text{$<$4:0$>$}}= P_{\text{$<$-3:0$>$}}=0.\end{align}
In the general SAV case, the integrablilty conditions for this formulation can be generated by cross-derivatives of Eq. 
\eqref{KILLPV2}. Before continuing, however, it is useful to define the following differential operators, which illuminates the structure in these equations. Let
\begin{align}
\mathcal{FP}(i,j)= 2 f_{{i,\zeta \zeta }} P_{{<j:4>}}+ 3  f_{{i,\zeta }} P_{{<j:4>,\zeta }} + f_i P_{{<j:4>,\zeta \zeta }}\notag\\
\overline{\mathcal{FP}}(i,j)= 2 f_{{i,\overline{\zeta} \overline{\zeta} }} P_{{<j:3>}}+ 3  f_{{i,\overline{\zeta} }} P_{{<j:3>,\overline{\zeta} }} + f_i P_{{<j:3>,\overline{\zeta} \overline{\zeta} }}.\label{FPOP}
\end{align}
The I.C.'s for the fields  $ \left\{P_{\text{$<$-1:0$>$}},P_{\text{$<$1:0$>$}},P_{\text{$<$5:0$>$}}\right\}$ can be written as 
\begin{align}
\mathcal{FP}(i,i)=\overline{\mathcal{FP}(i,i)}
\label{ICPv1}
\end{align}
where the index $i$ takes on the values $i\in\{1,2,4  \}$. Note that these ICs are equivalent to the ICs on the conformal factor of a two-metric in a two-dimensional spacetime that admits a fourth-order invariant \cite{JdB1}.  Thus in an abstract sense one is looking for at least three distinct interrelated two-metrics that admit fourth-order invariants in addition to satisfying  the ICs generated by the remaining fields. The ICs for the fields $\left\{P_{{<-2:0>}},P_{{<-3:0>}},P_{{<-4:0>}},P_{{<2:0>}},P_{{<4:0>}}\right\}$ take the form
\begin{align}
\mathcal{FP}(i,j) + \mathcal{FP}(j,i)=\overline{\mathcal{FP}(i,j)}+ \overline{\mathcal{FP}(j,i) },
\label{ICPv2}
\end{align}
where the index pair $(i,j)$ takes on the values  \mbox{$(i,j)=\{(1,2),(1,3),(1,4),(2,3),(3,4)\}$}.
The remaining IC for $P_{{<3:0>}}$ takes  the form
\begin{align}
2\mathcal{FP}(3,3) - &\mathcal{FP}(2,4) -\mathcal{FP} (4,2)=\notag\\ &\ \ 2 \overline{\mathcal{FP}(3,3)}-\overline{\mathcal{FP}(2,4)}- \overline{\mathcal{FP}(4,2) },
\label{ICPv3}
\end{align}
which slightly breaks the symmetry of the previous two sets of expressions \eqref{ICPv1} and  \eqref{ICPv2}. However, if in addition to the Killing equations, $f_{3}$ and $P_{<3:4>}$ obey the integrability conditions of a two-manifold with a fourth-order invariant, namely Eq. \eqref{ICPv1} with $i=3$, the symmetry is beautifully restored. Eq. \eqref{ICPv3} provides the missing pair $(i,j)= (2,4)$ in \eqref{ICPv2}.  The existence of a fourth-order invariant on a SAV spacetime then becomes synonmous with four interlocking two-manifolds admitting a fourth-order invariant (Eqs. \eqref{ICPv1}) with additional conditions (Eqs. \eqref{ICPv2}).

These ICs coupled with Eqs. \eqref{PKILL3}, differentiated a few more times can also be used to build a large linear system to perform  a brute force check. It does however turn out that they are of greater use as a framework for guessing the approximations of the invariant, whether or not it actually exists. A means of doing this systematically is explored in greater detail in \cite{JdBZV1} and  \cite{JdBZV2}.

\section{Conclusion} 
\label{SECCON}
In many ways the geodesics of a spacetime can be viewed as one of its  most fundamental descriptions.  The nature of the self-contained paths through space that massive particles  favor gives us a direct observational characterization of the inertial field and of  the  mass distribution within the spacetime.
The field equations of SAV spacetimes are completely integrable.
If the integrability properties are inherited by the geodesics of the manifold and  that relationship can be made explicit; the practical application of a nonperturbative description of geodesic motion in SAV spacetimes can possibly be implemented in an experimental environment.

This paper provides a framework by which the relationship between constants of motion, expressed as higher-order Killing tensors, and the curvature content of the spacetime can be quantified. To do so most elegantly it is suggested that the solution generation techniques themselves be exploited to simplify the calculation. This will  also increase our understanding of what these transformations imply physically  for the spacetimes they are mapping. If the ability to maintain the existence of a fourth-order Killing tensor for certain of the B\"{a}ckland transformations of a given spacetime can be quantified, it will give some insight into the size of the subgroup of SAV spacetimes that admit a fourth order Killing tensor.

The main sticking point at the moment is the the lack of a good method of eliminating all the Killing tensor components to obtain the conditions for the existence of the fourth-order Killing tensor on the $M$  variables without building up excessively large polynomials.
It is hoped that an elegant method of inverting the linear integrability matrix can be developed instead of the current ad hoc approach for choosing pivots. Failing that, however, the brute force method will eventually yield a result.

If all SAV spacetimes admit a fourth-order Killing tensor, then there is a direct relationship between the geodesic structure contained in the Killing equations and the algebraic structure of the Weyl tensor. It further implies that the pole structure of the Ernst equation can be understood in terms of the poles of the analytic functions hidden in the Killing equations. The field of two-degree-of-freedom dynamical systems, with its paucity of examples of systems admitting a fourth-order invariant will then  gain the solution generation techniques for Ernst's equations and the potential of a bi-infinite series of examples generated by the HKX transformations and other solution generation techniques. Moser presented a geometric picture of how integrability on the Jacobi ellipsoids arises, and of the relationship to quadrics, by means of a very simple geometric construction. Neugebauer \cite{Neu} pointed out that each axisymmetric stationary vacuum field corresponds to a minimal surface on a hyperbolic paraboloid $x+y+u^2+v^2-w^2=0$ embedded in a five-dimensional pseudo-Euclidean space. He further ventured that the B\"{a}ckland transformations were mapping minimal surfaces into minimal surfaces. It is possible that Moser's picture could be generalized to quartics which have their origin in the  Lorentz boosts and that using Neugebauer's observation a similar geometric picture could be built. This paragraph is speculative, it could list a lot of wishful thinking. In fact numerical evidence of chaotic behavior exists in some regions of phase space for the SAV spacetimes \cite{Gair}. This paper provides an algebraic test for the validity of these ideas. 

The alternative symmetric formulation of the Killing equations given in Sec.\ref{SecALT} has several advantages, despite its currently obscure mathematical origin. By confirming the hunch that the fourth-order Killing equations in 4D spacetime can be written in the form of four fourth-order Killing equations of a two-manifold with additional interlocking conditions, it gives greater insight into the structure of the Killing equations in 4D spacetimes. The high degree of symmetry of these equations and their compactness eases the way for further analytic exploration, increasing the possiblity for their analytic solution, which will be attempted in \cite{JdBZV1}. Furthermore, the simplifications that result from the additional assumption that  the spacetime is static become immediately apparent, eliminating a large number of variables from the problem. This fact, coupled with an additional assumption of equatorial symmetry, allows one to obtain an approximation for the Poincare map on the equatorial plane of these spacetimes. This program has been carried out for the Zipoy-Voorhees metric \cite{JdBZV2} and the results agree very well with numerical simulations of the orbits. Finding a more systematic derivation of the variables used in this formulation beyond just being the product of a fortuitous postulate should be an interesting exercise in itself.

When gravitational-wave observatories such as LIGO and LISA mature  and allow us to  probe the nether regions of spacetime around compact objects, we stand to learn a great deal.
 It is hoped that this and the preceding papers in this series, \cite{JdB0,JdB1,JdB2} will provide a mathematical framework in
 which to discuss practical algorithms for mapping  spacetimes. So doing to facilitate optimally decoding the information gleaned from gravitational wave observatories.   

\section{Acknowledgments}
My sincere thanks to Frank Estabrook for many useful discussions. I am also indebted to Tanja Hinderer and Michele Vallisneri for their insightful comments on the manuscript. I gratefully acknowledge support from NSF grants PHY-0653653, PHY-0601459, NASA grant NNX07AH06G, the Brinson Foundation and the David and Barbara Groce startup fund at Caltech.

\appendix

\begin{widetext}

\section{Fourth order Killing Equations in a Transverse Tetrad  }
 \label{APPENDTETRADIC}
Killing tensor components confined to the directions set by the SAV Killing vectors have their gradients fully defined by the Killing equations. These equations, expressed in terms of the directional derivatives and rotation coefficients are given below,
{\allowdisplaybreaks{ \begin{align}
T_{1111,3}&= 2 T_{1111} \gamma _{131}-4 T_{1112} \gamma _{131}-12 T_{1134} \gamma _{131}-4 T_{1111} \gamma _{132}-12 T_{1133} \gamma _{141}-2
   T_{1111} \gamma _{232},\notag\\
T_{{1111,4}}&= -12 T_{1144} \gamma _{131}+2 T_{1111} \gamma _{141}-4 T_{1112} \gamma _{141}-12 T_{1134} \gamma _{141}-4
   T_{1111} \gamma _{142}-2 T_{1111} \gamma _{242},\notag\\
T_{{2222,3}}&= -2 T_{2222} \gamma _{131}-4 T_{2222} \gamma _{132}-4 T_{1222} \gamma _{232}+2 T_{2222}
   \gamma _{232}-12 T_{2234} \gamma _{232}-12 T_{2233} \gamma _{242},\notag\\
T_{{2222,4}}&= -2 T_{2222} \gamma _{141}-4 T_{2222} \gamma _{142}-12 T_{2244} \gamma
   _{232}-4 T_{1222} \gamma _{242}+2 T_{2222} \gamma _{242}-12 T_{2234} \gamma _{242},\notag\\
T_{{1112,3}}&= \left(T_{1112} -3 T_{1122} -6
   T_{1234}\right) \gamma _{131}-\left(4 T_{1112} +6 T_{1134}\right) \gamma _{132}-6 T_{1233} \gamma _{141}-6 T_{1133} \gamma _{142}-\left(T_{1111} +T_{1112}\right)
   \gamma _{232},\notag\\
T_{{1112,4}}&= -6 T_{1244} \gamma _{131}-6 T_{1144} \gamma _{132}+\left(T_{1112} -3 T_{1122} -6 T_{1234}\right) \gamma
   _{141}-\left(4 T_{1112}+6 T_{1134}\right) \gamma _{142}-\left(T_{1111} +T_{1112}\right) \gamma _{242},\notag\\
T_{{1222,3}}&= -\left(T_{1222} +T_{2222}\right)
   \gamma _{131}- \left(4 T_{1222} +6 T_{2234}\right) \gamma _{132}-6 T_{2233} \gamma _{142}+\left(-3 T_{1122} +T_{1222} -6 T_{1234}\right) \gamma
   _{232}-6 T_{1233} \gamma _{242},\notag\\
T_{{1222,4}}&= -6 T_{2244} \gamma _{132}-\left(T_{1222} +T_{2222}\right) \gamma _{141}-\left(4 T_{1222} +6
   T_{2234}\right) \gamma _{142}-6 T_{1244} \gamma _{232}+\left(-3 T_{1122} +T_{1222} -6 T_{1234}\right) \gamma _{242},\notag\\
T_{{1122,3}}&= -2 (T_{1222}
  + T_{2234}) \gamma _{131}-4( T_{1122} +2T_{1234} )\gamma _{132}-2 T_{2233} \gamma _{141}-8 T_{1233} \gamma _{142}-2\left( T_{1112} + T_{1134}\right) \gamma _{232}-2 T_{1133} \gamma _{242},\notag\\
T_{{1122,4}}&= -2 T_{2244} \gamma _{131}-8 T_{1244} \gamma _{132}-2( T_{1222} +
   T_{2234}) \gamma _{141}-4 (T_{1122} +2 T_{1234}) \gamma _{142}-2 T_{1144} \gamma _{232}-2 \left(T_{1112}+ T_{1134}\right) \gamma _{242}. \label{KILLEQF1a}
\end{align}}}
The equations governing Killing tensor components with mixed indices $T_{A_1 A_2 k_1k_2 } $ that have fully defined gradients are,
{\allowdisplaybreaks{  \begin{align}
T_{{1134,3}}=& \frac{1}{2} \left(-T_{1133,4}-2 T_{1134}(2\gamma _{132}+ \gamma _{232}-\gamma_{131})- T_{1133} (2\gamma _{142}+ \gamma
   _{242}-\gamma_{141}+2 \gamma _{344})\right)\notag\\
&- ( 2 T_{1234}+ T_{3344}) \gamma _{131}
-(  T_{1233}+ T_{3334}) \gamma _{141},\notag\\
T_{{1134,4}}=& \frac{1}{2} \left(-T_{{1144,3}}-2 T_{1134}  \left( - \gamma _{141}+2  \gamma _{142}+ \gamma _{242}\right)+T_{1144} \left(\gamma
   _{131}-2 \gamma _{132}-\gamma _{232}+2 \gamma _{343}\right)\right)\notag\\
&-(T_{1244} +T_{3444}) \gamma
   _{131}-(2 T_{1234} +T_{3344}) \gamma _{141}, \notag\\
T_{{2234,3}}=&\frac{1}{2} \left(-T_{{2233,4}}-2 T_{2234} (\gamma_{131}+2
   \gamma _{132}-\gamma_{232 })-T_{2233} (\gamma _{141}+2  \gamma _{142}-\gamma _{242}+2  \gamma _{344})\right)\notag\\
&-(2 T_{1234} + T_{3344} ) \gamma _{232}-( T_{1233}+ T_{3334}) \gamma _{242},\notag\\
T_{{2234,4}}=& \frac{1}{2} \left(-T_{{2244,3}}-2
   \left( +T_{2234} \left(\gamma _{141}+2 \gamma _{142}-\gamma _{242}\right)
+\right)-T_{2244} \left(\gamma _{131}+2 \gamma _{132}-\gamma _{232}-2 \gamma _{343}\right)\right)\notag\\
&-(T_{1244} +T_{3444} )\gamma _{232}
-(2 T_{1234} +T_{3344})\gamma _{242},
\notag\\
T_{{1234,3}}=& \frac{1}{2}
   \left(-T_{{1233,4}}-2 T_{2234} \gamma _{131}-T_{2233} \gamma _{141}
-2 T_{1134} \gamma _{232}-T_{1133} \gamma _{242}-2 T_{1233} \gamma _{344}\right)\notag\\
&-(2 T_{1234} + T_{3344}) \gamma _{132}-( T_{1233} +
   T_{3334} )\gamma _{142},\notag\\
T_{{1234,4}}=& \frac{1}{2}
   \left(-T_{{1244,3}}-T_{2244} \gamma _{131}-2 T_{2234} \gamma _{141}
-T_{1144} \gamma _{232}-2 T_{1134} \gamma _{242}+2 T_{1244} \gamma _{343}\right)\notag\\
&-( T_{1244} + T_{3444}) \gamma _{132}-(2 T_{1234}+
   T_{3344}) \gamma _{142}. \label{KILLEQF2a}
\end{align}}}
Mixed components that have partially defind  gradients are governed by the equations;
{\allowdisplaybreaks{ \begin{align}
 T_{1133,3}&= T_{1133} \left(\gamma _{131}-2
   \gamma _{132}-\gamma _{232}-2 \gamma _{343}\right)-\frac{2}{3} \left(3 T_{1233} \gamma _{131}+T_{3334} \gamma _{131}+T_{3333} \gamma_{141}\right),\notag\\
T_{{2233,3}}&= -\frac{2}{3}  \left(3 T_{1233} \gamma _{232}+T_{3334} \gamma _{232}+T_{3333} \gamma _{242}\right)- T_{2233}
   \left(\gamma _{131}+2 \gamma _{132}-\gamma _{232}+2 \gamma _{343}\right),\notag\\
T_{{1233,3}}&= \frac{1}{3} \left(-3 T_{2233} \gamma _{131}-2 T_{3334}
   \gamma _{132}-2 T_{3333} \gamma _{142}-3 T_{1133} \gamma_{232}-6 T_{1233} \left(\gamma_{132}+\gamma_{343}\right)\right),\notag\\
T_{{1144,4}}&= -\frac{2}{3} \left( T_{4444} \gamma _{131}+ \left(3 T_{1244}+T_{3444}\right) \gamma_{141}\right)+ T_{1144} \left(\gamma_{141}-2 \gamma_{142}-\gamma_{242}+2 \gamma_{344}\right),\notag\\
T_{{2244,4}}&= -\frac{2}{3} \left( \left(T_{4444} \gamma_{232}+\left(3 T_{1244}+T_{3444}\right) \gamma _{242}\right) \right)- T_{2244}
   \left(\gamma_{141}+2 \gamma_{142}-\gamma _{242}-2 \gamma _{344}\right),\notag\\
T_{{1244,4}}&= \frac{1}{3} \left(-2 T_{4444} \gamma _{132}-3 T_{2244}
   \gamma _{141}-6 T_{1244} \gamma _{142}-2 T_{3444} \gamma _{142}-3 T_{1144} \gamma _{242}+6 T_{1244} \gamma _{344}\right). \label{KILLEQF2b}
\end{align}}}
The components orthogonal to the Killing directions $T_{(k_1k_2k_3k_4)}$,  $k_i=\{3,4\}$ constitute a subgroup of equations that represent the fourth order Killing equations of a two manifold. Within this subgroup components with gradients fully defined are
\begin{align}
T_{{3344,3}}&= -\frac{2}{3}
   \left(T_{{3334,4}}+2 T_{3334} \gamma _{344}\right),&T_{{3344,4}}&= -\frac{2}{3} \left(T_{{3444,3}}-2 T_{3444} \gamma_{343}\right). \label{KILLEQF3a}
\end{align}
The remaining components  with partially defined gradients are governed by the equations;
\begin{align}
T_{{3334,3}}=& -\frac{1}{4}T_{\text{3333,4}} - T_{3333} \gamma _{344}-2 T_{3334} \gamma
   _{343},&
T_{{3444,4}}=&  -\frac{1}{4}T_{\text{4444,3}}+ T_{4444} \gamma _{343} +2 T_{3444} \gamma _{344},\notag\\
T_{{3333,3}}=& -4 T_{3333}\ \gamma _{343},&
T_{{4444,4}}=& 4 T_{4444}\ \gamma _{344}. \label{KILLEQF3b}
\end{align}

\section{Integrability conditions  the general fourth-order case}
A large number of Killing Tensor components can be removed by writing down the integrability conditions (IC's) for all Killing tensor components whose gradients  have been fully defined, namely equations
\eqref{KILLEQF1a},\eqref{KILLEQF2a} and \eqref{KILLEQF3a}.
In particular the five tensor components $T_{1111},\ T_{1112},\ T_{1122},\ T_{1222},\ T_{2222}$ can be completely removed from consideration using  cross derivatives of Eqs. \eqref{KILLEQF1a} to yield the integrability conditions; 
{\allowdisplaybreaks{ \begin{align}
\left(  \begin{array}{l}
(6 T_{1244} +2 T_{3444}) \gamma _{131}^2+3 T_{{1144,3}} \gamma _{131}
+2T_{1144} \gamma _{{131,3}}
\\
+T_{1144} \left(-5 \gamma _{131}^2+\left(14 \gamma _{132}+5
   \gamma _{232}-4 \gamma _{343}\right) \gamma _{131}\right)
\end{array}\right)&
= \left(\begin{array}{l}
(6 T_{1233} +2 T_{3334}) \gamma _{141}^2+3 T_{{1133,4}}
   \gamma _{141}+2T_{1133} \gamma _{{141,4}} \\
-T_{1133} \left(5 \gamma _{141}^2-\left(14 \gamma _{142}+5 \gamma _{242}+4 \gamma _{344}\right) \gamma _{141}\right)\end{array}\right)
,\notag\\
\left( \begin{array}{l}
(6  T_{1244} +2 T_{3444}) \gamma _{232}^2+3 T_{{2244,3}} \gamma _{232}+2T_{2244} \gamma _{{232,3}}  \\
+T_{2244} \left((-5 \gamma _{232}+5 \gamma _{131} +14
   \gamma _{132} -4 \gamma _{343}) \gamma _{232}\right)\end{array}\right)&=
\left(\begin{array}{l}
(6 T_{1233} +2 T_{3334}) \gamma _{242}^2+3
   T_{{2233,4}} \gamma _{242}  +2T_{2233} \gamma _{{242,4}} \\
-T_{2233} \left((5 \gamma _{242}-5 \gamma _{141} -14 \gamma _{142} -4 \gamma _{344})\gamma_{242} \right) \end{array}\right),\notag\\
\left(\begin{array}{l}
+3 T_{{1244,3}} \gamma _{131}+3
   T_{{1144,3}} \gamma _{132} + 2 T_{1244} \gamma_{131,3}\\
+3 T_{2244} \gamma _{131}^2+4 T_{3444} \gamma _{132} \gamma _{131} +9T_{1144} \gamma _{131} \gamma _{232}  \\
+3T_{1144} \gamma _{132} \left(4 \gamma _{132}+\gamma _{232}-2 \gamma _{343}- \gamma _{131} \right)\\
+2 T_{1244} \left(-\gamma _{131}^2+\left(10 \gamma _{132}+\gamma _{232}-2 \gamma _{343}\right) \gamma _{131}\right)\end{array}\right) 
&=\left(\begin{array}{l}
+3 T_{{1233,4}} \gamma _{141}+3 T_{{1133,4}} \gamma
   _{142} + 2T_{1233  }\gamma
   _{{141,4}}  \\
+3 T_{2233} \gamma _{141}^2+4 T_{3334} \gamma _{142} \gamma _{141}  +9 T_{1133} \gamma _{141} \gamma _{242}  \\
+3 T_{1133} \left(\gamma _{142} \left(4 \gamma _{142}+\gamma _{242}-\gamma_{141}+2 \gamma
   _{344}\right)\right)\\
-2 T_{1233} \left(\gamma _{141}^2-\left(10 \gamma _{142}+\gamma _{242}+2 \gamma _{344}\right) \gamma _{141}\right)\end{array}  \right)
,\notag\\
\left(\begin{array}{l}
+3 T_{{1244,3}} \gamma _{232}+3 T_{{2244,3}}+2T_{1244}\gamma
   _{{232,3}} \gamma
   _{132}\\
+3 T_{1144} \gamma _{232}^2+4 T_{3444} \gamma _{132} \gamma _{232}+ 9 T_{2244} \gamma _{131} \gamma _{232}  \\
+3 T_{2244} \left(\gamma _{132} \left(\gamma_{131}+ 4 \gamma _{132}-\gamma _{232}-2 \gamma
   _{343}\right)\right)\\
+2 T_{1244} \left((-\gamma _{232}+\gamma _{131} +10 \gamma _{132} -2 \gamma _{343}) \gamma _{232}\right) \end{array}\right)
&=
\left(\begin{array}{l}
+3 T_{{1233,4}} \gamma _{242}+3 T_{{2233,4}} \gamma_{142}+2T_{1233}\gamma
   _{{242,4}}\\
+3 T_{1133} \gamma _{242}^2+4 T_{3334} \gamma _{142} \gamma _{242}
+9 T_{2233} \gamma _{141}  \gamma _{242}\\
+3 T_{2233} \left(\gamma _{142} \left(\gamma_{141}+ 4 \gamma _{142}-\gamma _{242}+2 \gamma
   _{344}\right)\right)\\
+2 T_{1233} \left((-\gamma _{242}+\gamma _{141} +10 \gamma _{142} +2 \gamma _{344}) \gamma _{242}\right) \end{array}
\right)
,\notag\\
\left(\begin{array}{l}
3 T_{{2244,3}} \gamma _{131}+12 T_{{1244,3}} \gamma _{132}+3 T_{{1144,3}} \gamma _{232}\\
+2 (T_{2244}\gamma_{131,3}+ T_{1144} \gamma_{232,3}) \\
+T_{3444} \left(8 \gamma
   _{132}^2+4 \gamma _{131} \gamma _{232}\right)\\
+T_{2244} \left(\gamma _{131}^2+\left(26 \gamma _{132}-\gamma _{232}-4 \gamma _{343}\right) \gamma _{131}\right)\\
+12 T_{1244} \left(4 \gamma _{132}^2-2 \gamma _{343} \gamma _{132}+3 \gamma _{131} \gamma _{232}\right)\\
+T_{1144} \left((\gamma
   _{232}-\gamma _{131} +26 \gamma _{132} -4 \gamma _{343}) \gamma _{232}\right) \end{array}\right)
&=\left(\begin{array}{l}3 T_{{2233,4}} \gamma
   _{141}+12 T_{{1233,4}} \gamma _{142}+3 T_{{1133,4}} \gamma _{242}\\
+2 (T_{2233}\gamma_{141,4}+ T_{1133} \gamma_{242,4}) \\
+4 T_{3334} \left(2 \gamma _{142}^2+\gamma _{141} \gamma _{242}\right)\\
+T_{2233}
   \left(\gamma _{141}^2+\left(26 \gamma _{142}-\gamma _{242}+4 \gamma _{344}\right) \gamma _{141}\right)\\
+12 T_{1233} \left(4 \gamma
   _{142}^2+2 \gamma _{344} \gamma _{142}+3 \gamma _{141} \gamma _{242}\right)\\
-T_{1133} \left((-\gamma _{242}+\gamma _{141} -26 \gamma _{142}
   -4 \gamma _{344}) \gamma _{242}\right)\end{array}\right).\label{IC1111}
\end{align}}}
Note that the five IC's of Eq. \eqref{IC1111} only contain first derivatives of the Killing tensor components.
The three tensor components $T_{1134},\ T_{2234},\ T_{1234}$ can be eliminated using  cross derivatives of Eqs. \eqref{KILLEQF2a} to yield the integrability conditions; 
{\allowdisplaybreaks{ 
\begin{align}
\left(
\begin{array}{l}
 T_{3444} \left(-6 \gamma _{131}^2+2 \left(12 \gamma _{132}+3 \gamma _{232}-7 \gamma _{343}\right) \gamma _{131}+6 \gamma _{\text{131,3}}\right)\\
+ 3
   T_{1144} \left(\gamma _{131}^2+\left(-4 \gamma _{132}+2 \gamma _{232}+3 \gamma _{343}\right) \gamma _{131}+6 \gamma _{132}^2+\gamma _{232}^2+2 \gamma
   _{343}^2-\gamma _{\text{131,3}} \right)\\
+ 3
   T_{1144} \left( 4 \gamma _{132} \gamma _{232}+\gamma _{\text{232,3}}-8 \gamma _{132} \gamma _{343}-3 \gamma _{232} \gamma _{343}-2 \gamma
   _{\text{343,3}}\right) \\
 -6 T_{1244} \left(\gamma _{131}^2-\left(4 \gamma _{132}+\gamma _{232}-3 \gamma _{343}\right) \gamma _{131}-\gamma
   _{\text{131,3}}\right) + 6 T_{2244} \gamma _{131}^2 \\
 10 T_{\text{3444,3}} \gamma _{131} + T_{\text{1144,3}} \left(-6 \gamma _{131}+12 \gamma _{132}+6 \gamma _{232}-9 \gamma _{343}\right) + 12
   T_{\text{1244,3}} \gamma _{131}  +  3 T_{\text{1144,33}} \\
\end{array}
\right)&=\left(\begin{array}{c}
\mbox{Interchange}\\
\mbox{ind. 3 and 4} \end{array}
\right),\notag\\
\left(
\begin{array}{l}
 2 T_{3444} \left(-3 \gamma _{232}^2+3 \gamma _{131} \gamma _{232}+12 \gamma _{132} \gamma _{232}-7 \gamma _{343} \gamma _{232}+3 \gamma
   _{\text{232,3}}\right) \\+ 3 T_{2244} \left(\gamma _{131}^2+\left(4 \gamma _{132}+2 \gamma _{232}-3 \gamma _{343}\right) \gamma
   _{131}+6 \gamma _{132}^2+\gamma _{232}^2+2 \gamma _{343}^2+\gamma _{\text{131,3}}\right)\\
 + 3 T_{2244} \left( -4 \gamma _{132} \gamma _{232}-\gamma _{\text{232,3}}-8 \gamma _{132}
   \gamma _{343}+3 \gamma _{232} \gamma _{343}-2 \gamma _{\text{343,3}}\right) \\
+  6 T_{1244} \left(-\gamma _{232}^2+\gamma _{131} \gamma _{232}+4 \gamma _{132} \gamma _{232}-3 \gamma
   _{343} \gamma _{232}+\gamma _{\text{232,3}}\right)+ 6 T_{1144} \gamma _{232}^2 \\
 + 10 T_{\text{3444,3}} \gamma _{232} + 12 T_{\text{1244,3}} \gamma _{232} + 3 T_{\text{2244,3}} \left(2 \gamma _{131}+4 \gamma _{132}-2 \gamma _{232}-3
   \gamma _{343}\right)
+ 3 T_{\text{2244,33}} \\
\end{array}
\right)&=\left(\begin{array}{c}
\mbox{Interchange}\\
\mbox{ind. 3 and 4} \end{array}\right),\notag\\
\left(
\begin{array}{l}
  2 T_{3444} \left(9 \gamma _{132}^2-10 \gamma _{343} \gamma _{132}+9 \gamma _{131} \gamma _{232}\right)\\+  3 T_{1144} \left(\gamma _{232}^2-\gamma _{131}
   \gamma _{232}+4 \gamma _{132} \gamma _{232}-3 \gamma _{343} \gamma _{232}+\gamma _{\text{232,3}}\right)\\+ 6 T_{1244} \left(3 \gamma _{132}^2-4 \gamma _{343}
   \gamma _{132}+\gamma _{343}^2+3 \gamma _{131} \gamma _{232}-\gamma _{\text{343,3}}\right) \\+ 3 T_{2244} \left(\gamma _{131}^2+\left(4 \gamma _{132}-\gamma
   _{232}-3 \gamma _{343}\right) \gamma _{131}+\gamma _{\text{131,3}}\right) \\
+ 10 T_{\text{3444,3}} \gamma _{132} + 6 T_{\text{1144,3}} \gamma _{232} + 3 T_{\text{1244,3}} \left(4 \gamma _{132}-3 \gamma _{343}\right) + 6
   T_{\text{2244,3}} \gamma _{131} +
 T_{\text{1244,33}} 
\end{array}
\right)&=\left(\begin{array}{c}
\mbox{Interchange}\\
\mbox{ind. 3 and 4} \end{array}\right).
\label{IC1133}
\end{align}
The structure of the three  ICs of Eq. \eqref{IC1133}  differ slightly from that of Eqs. \eqref{IC1111} in that they contain second derivatives of the Killing tensor components.
Finally the component $T_{3344}$ can be eliminated using  cross derivatives of Eqs. \eqref{KILLEQF3a} to yield the IC; 
\begin{align}
2 T_{3444} \left(\gamma _{343}^2-\gamma _{\text{343,3}}\right)  -3 T_{\text{3444,3}} \gamma _{343} +  T_{\text{3444,33}} 
=2 T_{3334} \left(\gamma _{344}^2+\gamma _{\text{344,4}}\right) + 3 T_{\text{3334,4}} \gamma _{344} +  T_{\text{3334,44}}\label{IC3344}
\end{align}}}
Eqs. \eqref{IC1111}, \eqref{IC1133} and \eqref{IC3344} constiture the nine basic integrability conditions on the remaining fields used to build up the linear system of equations for the Killing tensor components and their derivatives. 

\end{widetext}

\bibliographystyle{apsrev}
\bibliography{../BholesNemadon}

\begin{thebibliography}{15}
\expandafter\ifx\csname natexlab\endcsname\relax\def\natexlab#1{#1}\fi
\expandafter\ifx\csname bibnamefont\endcsname\relax
  \def\bibnamefont#1{#1}\fi
\expandafter\ifx\csname bibfnamefont\endcsname\relax
  \def\bibfnamefont#1{#1}\fi
\expandafter\ifx\csname citenamefont\endcsname\relax
  \def\citenamefont#1{#1}\fi
\expandafter\ifx\csname url\endcsname\relax
  \def\url#1{\texttt{#1}}\fi
\expandafter\ifx\csname urlprefix\endcsname\relax\def\urlprefix{URL }\fi
\providecommand{\bibinfo}[2]{#2}
\providecommand{\eprint}[2][]{\url{#2}}

\bibitem[{\citenamefont{Brink}(2008{\natexlab{a}})}]{JdB0}
\bibinfo{author}{\bibfnamefont{J.}~\bibnamefont{Brink}},
  \bibinfo{journal}{Phys.\ Rev.\ D} \textbf{\bibinfo{volume}{78}},
  \bibinfo{eid}{102001} (\bibinfo{year}{2008}{\natexlab{a}}).

\bibitem[{\citenamefont{Brink}(2008{\natexlab{b}})}]{JdB1}
\bibinfo{author}{\bibfnamefont{J.}~\bibnamefont{Brink}},
  \bibinfo{journal}{Phys.\ Rev.\ D} \textbf{\bibinfo{volume}{78}},
  \bibinfo{eid}{102002} (\bibinfo{year}{2008}{\natexlab{b}}).

\bibitem[{\citenamefont{Brink}(2009{\natexlab{a}})}]{JdB2}
\bibinfo{author}{\bibfnamefont{J.}~\bibnamefont{Brink}}, \bibinfo{journal}{III
  (in preparation), Second order Killing Tensors}
  (\bibinfo{year}{2009}{\natexlab{a}}).

\bibitem[{\citenamefont{Walker and Penrose}(1970)}]{Walker}
\bibinfo{author}{\bibfnamefont{M.}~\bibnamefont{Walker}} \bibnamefont{and}
  \bibinfo{author}{\bibfnamefont{R.}~\bibnamefont{Penrose}},
  \bibinfo{journal}{Commun. Math. Phys} \textbf{\bibinfo{volume}{18}},
  \bibinfo{pages}{265} (\bibinfo{year}{1970}).

\bibitem[{\citenamefont{Chandrasekhar}(1983)}]{MathTheoryofBlackHoles}
\bibinfo{author}{\bibfnamefont{S.}~\bibnamefont{Chandrasekhar}},
  \emph{\bibinfo{title}{The Mathematical Theory of Black Holes}}
  (\bibinfo{publisher}{Clarendon Press. Oxford}, \bibinfo{year}{1983}).

\bibitem[{\citenamefont{Gair et~al.}(2008)\citenamefont{Gair, Li, and
  Mandel}}]{Gair}
\bibinfo{author}{\bibfnamefont{J.~R.} \bibnamefont{Gair}},
  \bibinfo{author}{\bibfnamefont{C.}~\bibnamefont{Li}}, \bibnamefont{and}
  \bibinfo{author}{\bibfnamefont{I.}~\bibnamefont{Mandel}},
  \bibinfo{journal}{Phys.\ Rev.\ D} \textbf{\bibinfo{volume}{77}},
  \bibinfo{pages}{024035} (\bibinfo{year}{2008}).

\bibitem[{\citenamefont{Stephani et~al.}(2003)\citenamefont{Stephani, Kramer,
  Maccallum, Hoenselaers, and Herlt}}]{ExactSolutions}
\bibinfo{author}{\bibfnamefont{H.}~\bibnamefont{Stephani}},
  \bibinfo{author}{\bibfnamefont{D.}~\bibnamefont{Kramer}},
  \bibinfo{author}{\bibfnamefont{M.}~\bibnamefont{Maccallum}},
  \bibinfo{author}{\bibfnamefont{C.}~\bibnamefont{Hoenselaers}},
  \bibnamefont{and} \bibinfo{author}{\bibfnamefont{E.}~\bibnamefont{Herlt}},
  \emph{\bibinfo{title}{Exact Solutions of Einstein's Field Equations}}
  (\bibinfo{publisher}{Cambridge University Press}, \bibinfo{year}{2003}),
  \bibinfo{edition}{2nd} ed.

\bibitem[{\citenamefont{Wolf}(1998)}]{WolfT}
\bibinfo{author}{\bibfnamefont{T.}~\bibnamefont{Wolf}}, \bibinfo{journal}{Comp.
  Phys. Comm} pp. \bibinfo{pages}{316--329} (\bibinfo{year}{1998}).

\bibitem[{\citenamefont{Harrison}(1983)}]{HarrisonWET}
\bibinfo{author}{\bibfnamefont{B.~K.} \bibnamefont{Harrison}},
  \bibinfo{journal}{J. Math. Phys.} \textbf{\bibinfo{volume}{24}},
  \bibinfo{pages}{2178} (\bibinfo{year}{1983}).

\bibitem[{\citenamefont{Neugebauer}(1979)}]{Neu}
\bibinfo{author}{\bibfnamefont{G.}~\bibnamefont{Neugebauer}},
  \bibinfo{journal}{J. Phys. A: Math. Gen} \textbf{\bibinfo{volume}{12}},
  \bibinfo{pages}{L67} (\bibinfo{year}{1979}).

\bibitem[{\citenamefont{Hall}(1983)}]{Hall}
\bibinfo{author}{\bibfnamefont{L.~S.} \bibnamefont{Hall}},
  \bibinfo{journal}{Physica D} \textbf{\bibinfo{volume}{8}},
  \bibinfo{pages}{90} (\bibinfo{year}{1983}).

\bibitem[{\citenamefont{Brink}(2009{\natexlab{b}})}]{JdB3}
\bibinfo{author}{\bibfnamefont{J.}~\bibnamefont{Brink}}, \bibinfo{journal}{IV
  (in preparation), Relationship between Weyl Curvature and Killing Tensors in
  SAV Spacetimes}  (\bibinfo{year}{2009}{\natexlab{b}}).

\bibitem[{\citenamefont{Klein and Richter}(2005)}]{ErnstEQ}
\bibinfo{author}{\bibfnamefont{C.}~\bibnamefont{Klein}} \bibnamefont{and}
  \bibinfo{author}{\bibfnamefont{O.}~\bibnamefont{Richter}},
  \emph{\bibinfo{title}{Ernst Equation and Riemann Surfaces}}
  (\bibinfo{publisher}{Springer-Verlag Berlin Heidelberg},
  \bibinfo{year}{2005}).

\bibitem[{\citenamefont{Brink}(2009{\natexlab{c}})}]{JdBZV2}
\bibinfo{author}{\bibfnamefont{J.}~\bibnamefont{Brink}},
  \bibinfo{journal}{Poincare Maps of Static Spacetimes with Equatorial
  Symmetry-Example Zipoy Voorhees Metric.}
  (\bibinfo{year}{2009}{\natexlab{c}}).

\bibitem[{\citenamefont{Brink}(2009{\natexlab{d}})}]{JdBZV1}
\bibinfo{author}{\bibfnamefont{J.}~\bibnamefont{Brink}},
  \bibinfo{journal}{Formal solution of the Fourth Order Killing tensors
  equations of SAV Spacetimes.}  (\bibinfo{year}{2009}{\natexlab{d}}).

\end{thebibliography}

\end{document}